\documentclass[vecphys,multphys,sechang]{svmult}

\usepackage{ergebnisse}
\usepackage[authoryear]{natbib}

\usepackage{makeidx}         
\usepackage{graphicx}        
                             
\usepackage{multicol}        
\usepackage[bottom]{footmisc}

\makeindex

\begin{document}


\title{Formation, Dynamical Evolution, and Habitability of Planets
in Binary Star Systems}
\titlerunning{Extrasolar Planets in Binary Star Systems} 
\author{Nader Haghighipour}
\maketitle

\pagenumbering{arabic}

\begin{abstract}

A survey of currently known planet-hosting stars indicates that
approximately 25\% of extrasolar planetary systems are within dual-star
environments. Several of these systems contain stellar companions
on moderately close orbits, implying that studies of the formation and
dynamical evolution of giant and terrestrial planets, 
in and around binary star systems have now found realistic grounds.
With the recent launch of the space telescope CoRoT, 
and the launch of NASA's Kepler satellite in 2009, the number 
of such dynamically complex systems will soon
increase and many more of their diverse and  interesting
dynamical characteristics will soon be discovered.
It is therefore, both timely and necessary, 
to obtain a deep understanding of the history and current status of 
research 
on planets in binary star systems. This chapter will serve this purpose 
by reviewing the models of the formation of giant and 
terrestrial planets in dual-star environments, and by presenting results
of the studies of their dynamical evolution and 
habitability, as well as the mechanisms 
of delivery of water and other volatiles to their terrestrial-class
objects. In this chapter, the reader is presented with a
comprehensive, yet relatively less technical approach to
the study of planets 
in and around binary stars, and with discussions on 
the differences between dynamical 
characteristics of these systems and planetary systems around single stars.
\end{abstract}

\section{Introduction}
\label{sec:1}

The concept of a ``world with two suns'' has been of interest to
astronomers for many years. Many scientists tried to understand
whether planets could form in binary star systems, and whether
the notion of habitability, as we know it, could be extended
to such environments. Although as a result of their respective works, 
many dynamical features of {\it binary-planetary} systems\footnote{
A binary-planetary system is a dual-star system that also hosts planetary
bodies.} have been 
discovered, until recently,
the subjects of their studies were, in large part, hypothetical. There was
no detection of a planet in and/or around a binary system, 
and planet detection 
techniques had not 
advanced enough to successfully
detect planets in dual-star environments.

\begin{figure}
\centering
\includegraphics[height=7cm]{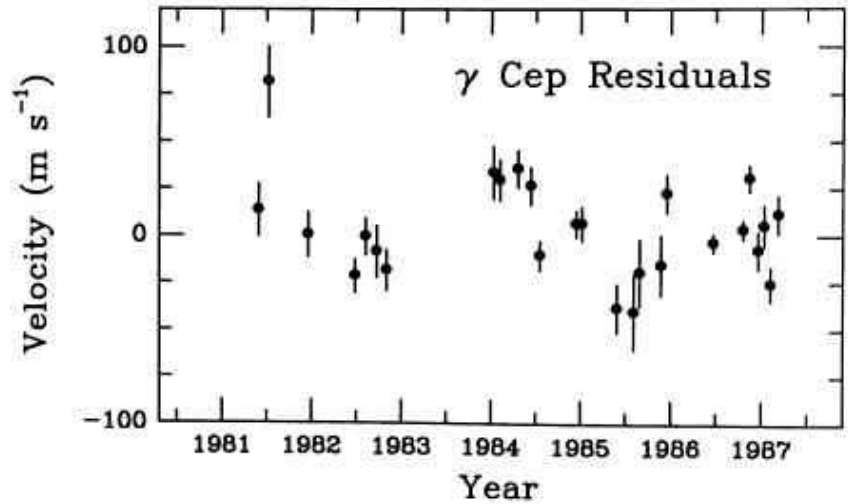}
\caption{Velocity residuals to $\gamma$ Cephei 
after subtracting a second order fit to its original radial velocity data 
\citep{Campbell88}. The residuals show periodicity implying the possible
existence of a planetary companion.}
\label{fig:1}
\end{figure}

The discovery of extrasolar planets during the past decade has, however, 
changed this trend. Although the candidate planet-hosting stars have 
been routinely chosen to be single, or within wide ($>$100 AU)
binaries \footnote{As shown by \citet{Jim02},  
the perturbative effect of the stellar companion on the 
dynamics of a planetary system around a star becomes important 
when the binary separation is smaller than 100 AU. At the present, 
approximately 25\% of extrasolar planetary systems detected by radial 
velocity technique are in binary systems with separations ranging from 
250 to 6000 AU.}, the precision radial velocity technique has been 
successful in detecting planets around the primaries of three moderately 
close ($<$40 AU) dual-star systems. As a result, during the past 
few years, the topic 
of planets in binaries, once again, found its way to the mainstream 
research and has now become a real scientific issue that demands 
theoretical explanations.

The first detection of a planet in a binary system was reported
by Campbell, Walker \& Yang in 1988. 
In an attempt to identify planetary objects outside 
our solar 
system, these authors measured the variations in the radial
velocities of a number of stars, and reported the possibility of 
the presence
of a Jovian-type body around the star $\gamma$ Cephei
\citep[figure 9.1,][]{Campbell88}. This star, that is 
a K1 IV sub-giant with a mass of 1.59 solar-masses \citep{Fuhr03}, is the
primary of a binary system with a semimajor axis of 18.5 AU 
and an eccentricity of 0.36 \citep{Griffin02,Hatzes03}. 
The secondary of this system  
is an M dwarf with a mass of 0.44 solar-masses 
\citep{Neuhauser07,Torres07}. Initial radial 
velocity
measurements of $\gamma$ Cephei implied that this star may be host to a 
giant planet with a probable mass of 1.7 Jupiter-masses, in
an orbit with a semimajor axis of 1.94 AU \citep{Campbell88}.

Unfortunately, the discovery of the first binary-planetary system,
which could have also marked the detection of the first planet 
outside our solar system, did not withstand skepticism. In an article 
in 1992, Walker and his colleagues
attributed their measured variations of the radial velocity of 
$\gamma$ Cephei to the chromospheric activities of this star,
and announced that the possibility of the existence of a giant planet
around $\gamma$ Cephei may be none \citep{Walker92}. 
It took observers an additional 12 years  
to monitor  $\gamma$ Cephei and measure its radial velocity 
to arrive at the conclusion that the previously observed variations 
were not due to stellar activities and were in fact representative of 
a planetary companion (figure 9.2). It was the
discovery of a giant planet in the binary system of GL 86 \citep{Queloz00}, 
and the (re-)announcement of the detection of a giant planet in 
$\gamma$ Cephei system \citep{Hatzes03} that opened a new chapter in 
the theoretical and observational studies of extrasolar planetary systems.

\begin{figure}
\centering
\vskip 0.2in
\includegraphics[height=7.3cm]{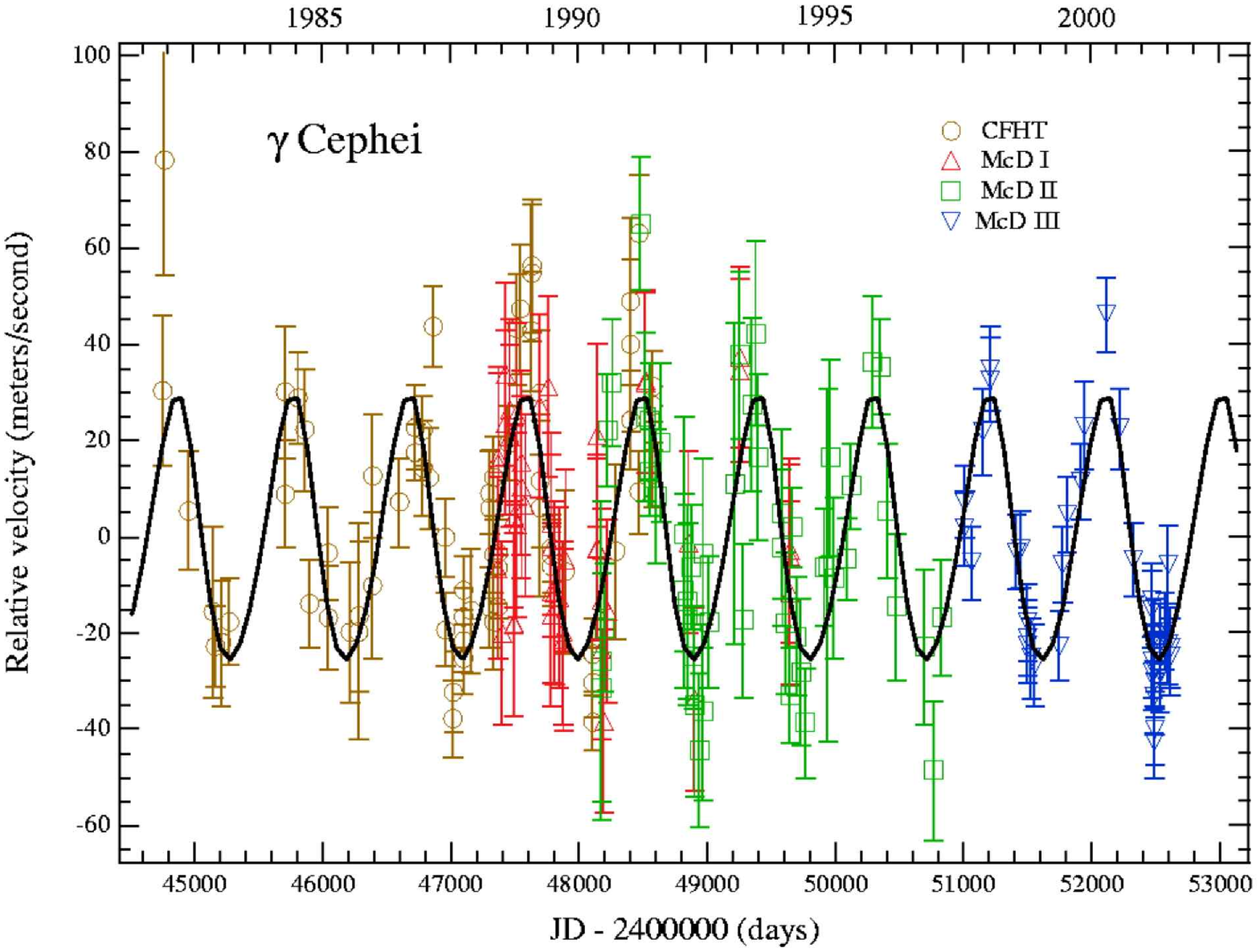}
\vskip 0.2in
\includegraphics[height=8.3cm]{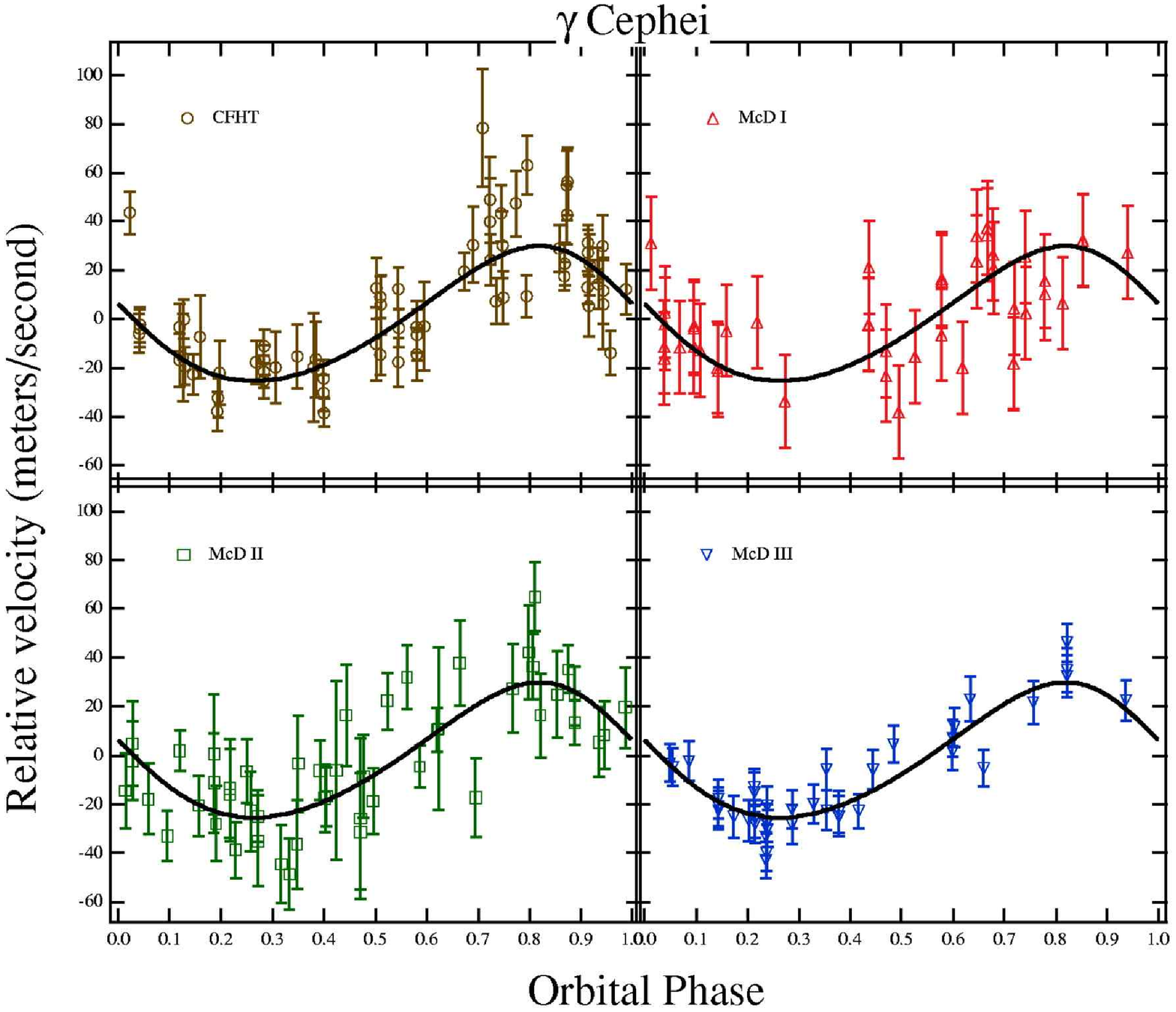}
\caption{Radial velocity measurements of the primary of 
$\gamma$ Cephei binary system. The graph on the top shows the
orbital solution for the planet (solid line) and the residual 
velocity measurements after subtracting the 
contribution due to the binary companion (data points). The
graph on the bottom depicts the phased residual radial velocity
measurements (data points) compared to the planet orbital solution 
(solid line) \citep{Hatzes03}.}
\label{fig:2}
\end{figure}

The fact that giant planets exist in moderately close 
($<40$ AU) binary systems has confronted dynamicists
with many new challenges. Questions such as, how are 
these planets formed, can binary-planetary systems host
terrestrial and/or habitable planets, how are habitable
planets formed in such dynamically complex environments, and
how do such planets acquire the ingredients necessary for
life, are among major topics of research in this area. This chapter
is devoted to review these issues and present the current status of
research on the formation of planet in dual-star systems and habitability 
of terrestrial bodies in and around binary stars. 

The chapter begins with a review of
the dynamics of a planet in a binary star system.
In general, prior to constructing a theory for the formation of 
planets, it proves useful to study whether the orbit of a planet
around its host star would be stable. In a binary system,
such studies are of quite importance since in these systems 
the perturbation of the stellar companion may dictate the
possibility of the formation of planetary bodies
by affecting the stability and dynamics of smaller objects.

The formation of planets in binary star systems is reviewed 
in the third section.
Although the study of the dynamics of planets in and around binary 
stars dates back to approximately forty year ago, the formation of 
planets in these systems is an issue that is still unresolved. 
In spite the 
observational evidence that indicate majority of main and pre-main
sequence stars are formed in binaries 
or clusters \citep{Abt79,Duq91,Math94,Math00,White01}, 
and in spite the detection of 
potentially planet-forming environments in and around binary stars 
\citep[figure 9.3, also see][]{Math94,Akeson98,Rodriguez98,White99,
Silbert00,Math00}, 
planet formation theories are still unclear in explaining 
how planets may form
in multi-star environments. The focus of section 9.3 is on discussing
the formation of giant and terrestrial planets in moderately close 
binary-planetary systems, and reviewing the current status of planet
formation theories in this area.

The habitability of a binary system is presented in section 9.4.
The notion of habitability is defined based on the habitability
of Earth and life, as we know it. Such a definition 
requires a habitable planet to 
have the capability of retaining liquid water in its atmosphere and
on its surface. 
The latter is determined by the luminosity of the central star,
the size of the planet, and also the distribution of water in the 
protoplanetary disk 
from which terrestrial-class objects are formed. In section 9.4, a
review of the current status of the models of habitable planet
formation in and around binary systems
are presented, and their connections to 
\hfill
\vskip 1pt
\begin{figure}
\centering
\vskip 0.15in
\includegraphics[height=10cm]{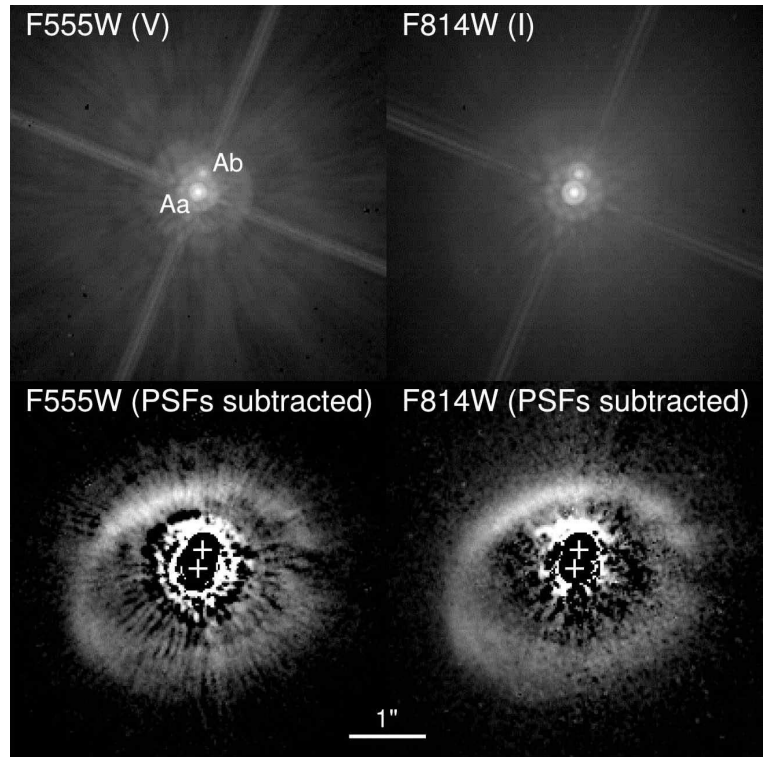}
\vskip 5pt
\caption{Image of the circumbinary disk of GG Tau \citep{Krist05}
taken by Advanced Camera for Surveys (ACS) on board of
Hubble Space Telescope (HST).
The binary separation is 35 AU.
The locations of the binary components are marked with crosses.} 
\label{fig:3}
\end{figure}
\vskip 5pt
\noindent
models of terrestrial planet formation and water-delivery
around single stars are
discussed. Finally the chapter ends 
by discussing the future prospects of research in the field of
planets in binaries.

\section{Dynamical Evolution and Stability}
\label{sec:1}

In general, one can consider orbital stability synonymous with  
the capability of an object in maintaining its orbital parameters
(i.e., semimajor axis, eccentricity, and inclination) at all times.
In other words, an object is stable if small variations in its orbital
parameters do not progress exponentially, but instead vary sinusoidally. 
Instability occurs when the perturbative forces create drastic
changes in the time variations of these parameters and result in 
either the ejection of the object from the system (i.e., leaving 
the system's gravitational field), or its collision with other bodies.

\begin{figure}
\centering
\vskip 0.2in
\includegraphics[height=4cm]{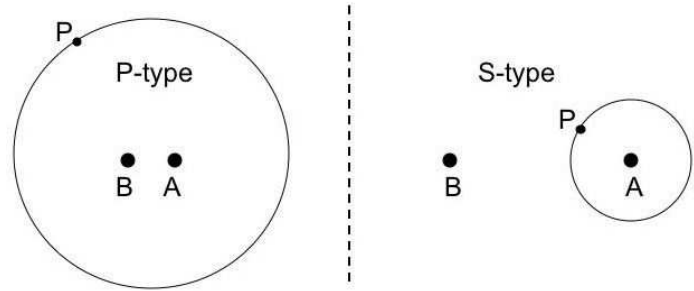}
\caption{S-type and P-type binary-planetary systems. A and B represent the
stars of the binary, and P depicts the planet.}
\label{fig:4}
\end{figure}

The concept of stability, as explained above, although simple, has been 
defined differently by different authors.
 A review by Szebehely in 1984 lists 50 different definitions
for the stability of an object in a multi-body system \citep{Szebehely84}.
For instance, \citet{Har77} considered the 
orbit of an object stable if, while numerically integrating the object's
orbit, its semimajor axis and orbital eccentricity would not undergo
secular changes. \citet{Szeb80}, and \citet{Szebehely81}, 
on the other hand, considered
integrals of motion and curves of zero velocity to determine the
stability of a planet in and around binary stars. In this chapter, 
the values
of the orbital eccentricity of an object and its semimajor axis 
are used to evaluate its stability. An object is considered stable if, for
the duration of the integration of its orbit, the value of its orbital 
eccentricity 
stays below unity, it does not collide with other bodies, and does not
leave the gravitational field of its host star.

The study of the stability of a planetary orbit in dual stars 
requires a detailed analysis of the dynamical evolution
of a three-body system. Such an analysis itself is dependent upon the
type of the planetary orbit.
In general, a planetary-class object may have three types of orbit
in and around a binary star system. \citet{Szeb80} and \citet{Dvorak83}
have divided these orbits into three different categories.
As indicated by \citet{Szeb80}, a planet may be in an {\it inner} orbit,
where it revolves around the primary star, or it may be in a 
{\it satellite} orbit, where it revolves around the secondary star.
A planet may also be revolving the entire binary system
in which case its orbit is called an {\it outer} orbit.
As classified by \citet{Dvorak83}, on the
other hand, a study of the stability of resonant periodic orbits in a
restricted, circular, three-body system indicates that a planet may 
have an S-type orbit, where it revolves around only one of the stars 
of the binary, or may be in a P-type orbit, where it revolves the entire
binary system (figure 9.4). A planet may also be in an L-type orbit where it
librates in a stable orbit around the $L_4$ or $L_5$ Lagrangian points.

The rest of this section is devoted to a review of the studies of
the stability of planets in S-type and P-type orbits. 
Given that in a binary star system, a planet
is subject to the gravitational attraction of two massive bodies 
(i.e., the stars), it would be important to understand how the
process of the formation of planetary objects, and the
orbital dynamics of small bodies would be affected by the orbital
characteristics of the binary's stellar components.
In general, except for some special cases for which analytical 
solutions may exist, such studies
require numerical integrations of the orbits
of all the bodies in the system. In the past, prior to the
invention of symplectic integrators \citep{Wisdom91}, which enabled dynamicists
to extend the studies of the stability of planetary systems to
several hundred million years, and before the development of fast
computers, majority of such studies were either limited to those 
special cases, 
or were carried out numerically for only a small number of 
binary's orbital period.
Examples of such studies can be found in articles by
\citet{Black81}, \citet{Black82}, and \citet{Black83}, in which the
authors studied the orbital stability of a planet
in and around a binary star. By numerically integrating
the equations of the motion of the planet, this authors showed that, 
when the stars of the binary have equal masses,
the orbital stability of the planet is independent of its
orbital inclination \citep[also see][]{Har77}. Their integrations also 
indicate that, 
when the mass of one of the stellar components is comparable
to the mass of Jupiter, planetary orbits with inclinations higher than 
50$^\circ$ tend to become unstable.

The invention of symplectic integrators, in particular routines that have
been designed  specifically for the purpose of integrating orbits 
of small bodies in dual-star systems\footnote{Symplectic integrators, 
as they were
originally developed by \citet{Wisdom91}, are not suitable
for numerically integrating the orbits of small bodies in the
gravitational fields of two massive objects. These integrators
have been designed to integrate the orbits of planetary or
smaller objects when they revolve around only one massive central
body. Recently \citet{Chambers02} have developed a version of a 
symplectic integrator that is capable of integrating the motion
of a small object in the gravitational fields of two stellar bodies.}
\citep{Chambers02}, have now enabled dynamicists to extend studies of planet
formation and stability in dual-star environments to much 
larger timescales. In the following, the results of such studies are
discussed in more detail.

\subsection {Stability of S-type orbits}

As mentioned above, instability occurs when the perturbative effects
cause the semimajor axis and orbital eccentricity of a planet 
change in such a way that either the object leaves the gravitational 
field of the system, or it collides with another body. For a planet
in an S-type orbit, the gravitational force of the secondary star
is the source of these perturbations. That implies, a planet at a
large distance from the secondary, i.e., in an
orbit closer to its host star, may receive less perturbation from the
binary companion and may be able to sustain its dynamical state for a
longer time \citep{Har77}. Since 
the perturbative effect of the stellar companion 
varies with  its mass, 
and the eccentricity and semimajor axis of the binary (which together
determine the closest approach of the secondary to the planet),
it is pos-

\begin{figure}
\centering
\includegraphics[height=6cm]{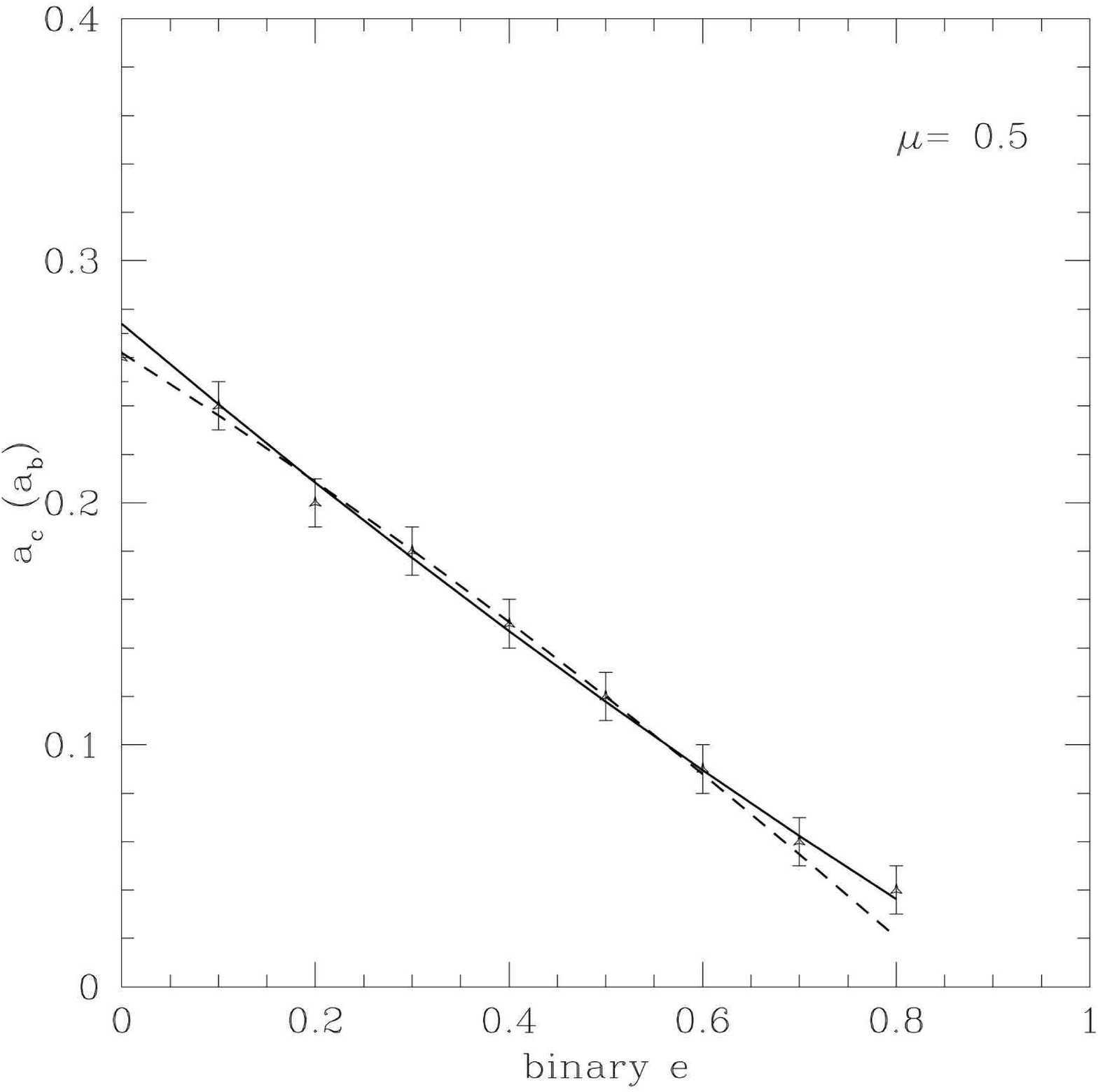}
\includegraphics[height=6cm]{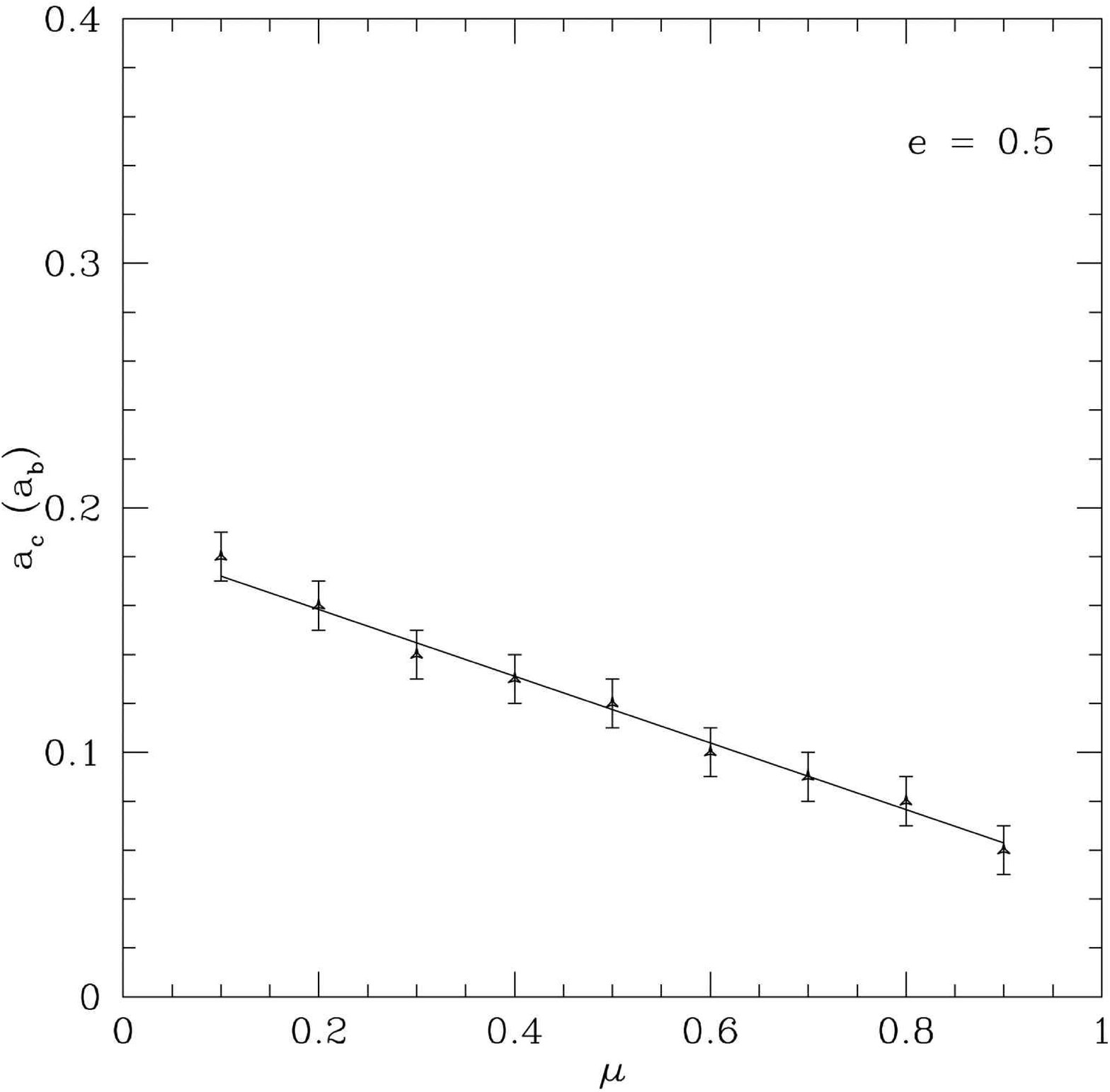}
\caption{Graphs of the critical semimajor axis $(a_c)$
of an S-type binary-planetary system, in units of 
the binary semimajor axis \citep{Holman99}. The graph on the left 
shows $a_c$ as a function of the binary eccentricity for an equal-mass 
binary. The graph on the right corresponds to the variations of
the critical semimajor axis of a binary with an eccentricity of
0.5 in term of the binary's mass-ratio. The solid and dashed line on
the left panel depict the empirical formulae as reported by
\citet{Holman99} and \citet{Dvorak88a}, respectively.}
\label{fig:5}
\end{figure}

\noindent
sible to estimate an upper limit for the planet's distance
to the star beyond which
the orbit of the planet would be unstable.
As shown by \citet{Dvorak88a} and
\citet{Holman99}, the maximum
value that the semimajor axis of a planet in an S-type orbit
can attain and still 
maintain its orbital stability 
is a function of the mass-ratio 
and orbital elements of the binary, and is given by 
\citep{Dvorak88a,Holman99}

\begin{eqnarray}
&\!\!\!\!\!\!\!\!\!\!\!\!\!\!\!\!\!\!\!\!\!\!\!\!
{{a_c}/{a_b}}=(0.464\pm 0.006)+ (-0.380 \pm 0.010)\mu 
+ (-0.631\pm0.034) {e_b} \nonumber \\
&\qquad+(0.586 \pm 0.061) \mu {e_b}
+ (0.150 \pm 0.041) {e_b^2} 
+(-0.198 \pm 0.047)\mu {e_b^2}\>.
\end{eqnarray}
\noindent
In this equation, $a_c$ is 
{\it critical} semimajor axis , $\mu={{M_1}/{({M_1}+{M_2}})}$, $a_b$ and $e_b$
are the semimajor axis and eccentricity of the binary, and 
$M_1$ and $M_2$ are the masses of the primary and secondary stars,
respectively. The $\pm$ signs in equation (9.1) define a lower
and an upper value for the critical semimajor axis $a_c$, and set
a transitional region that consists of a mix of 
stable and unstable systems. 
Such a dynamically {\it gray} area, in which the
state of a system changes from stability to instability, 
is known to exist in multi-body environments, and is a characteristic 
of any dynamical system.

Equation (9.1) is an empirical formula that has been obtained by
numerically integrating the orbit of a test particle (i.e., a
massless object) at different distances from the primary of a binary star
\citep{Dvorak88a,Holman99}. Figure 9.5 shows this in more detail.
Similar studies have been done
by \citet{Moriwaki04}, and \citet{Fatuzzo06} 
\begin{figure}
\centering
\includegraphics[height=5cm]{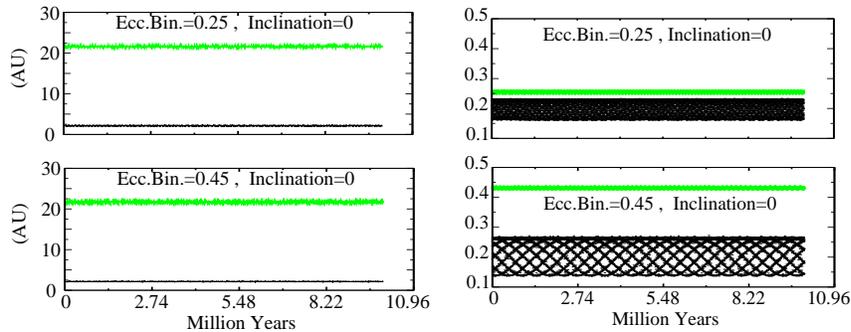}
\caption{Graphs of the semimajor axes (left) and eccentricities (right) 
of the giant planet (black) and binary (green) of $\gamma$ Caphei
for different values of the eccentricities of the binary \citep{Hagh03}.
The mass-ratio of the binary is 0.2.}
\label{fig:6}
\end{figure}
\noindent
who obtained
critical semimajor axes slightly larger than given by
equation (9.1).

Since the mass of a Jovian-type planet is approximately three
orders of magnitude smaller than the mass of a star, such a
test particle approximation yields 
results that not only are
applicable to the stability of giant planets, but
can also be used in identifying regions where
smaller bodies, such as terrestrial-class objects 
\citep{Quintana02,Quintana06,Quintana07} 
and dust particles \citep{Trilling07},
can have long term stable orbits\footnote{In applying equation (9.1)
to the stability of dust particles, one has to note that
this equation does not take into account the effects of 
non-gravitational forces such as gas-drag or radiation pressure.
The motion of a dust particles can be strongly
altered by the effects of these forces.}.
In a recent article, \citet{Trilling07} utilized equation (9.1)
and its stability criteria to explain the dynamics of debris
disks, and the possibility of the formation and 
existence of planetesimals in and around 22 binary star systems. 
By detecting infrared excess of dust particles, these authors 
confirmed the presence of stable dust bands, possibly 
resulted from collision 
of planetesimals, in S-type orbits in several wide binaries.

The stability of S-type systems has been studied by many authors
\citep{Benest88,Benest89,Benest93,Benest96,Holman97,
Pilat02,Dvorak03,Dvorak04,Pilat04,Musielak05}. In a recent article,
\citet{Hagh06} extended such studied to the dynamical stability 
of the Jupiter-like planet of the $\gamma$ Cephei planetary system.
By numerically integrating the orbit of this object for different 
values of ${a_b}, \, {e_b}$ and $i_p$ (the orbital inclination of
the giant planet relative to the plane of the binary), \citet{Hagh06} 
has shown that the orbit of this planet is stable for the values
of the binary eccentricity within the range $0.2 \leq {e_b} \leq 0.45$.
Figure 9.6 shows the results of such integrations for a coplanar system 
with $\mu=0.2$ and for different values of the binary  eccentricity.
The initial value of the semimajor axis of the binary was chosen to be

\begin{figure}
\centering
\includegraphics[height=3.5cm]{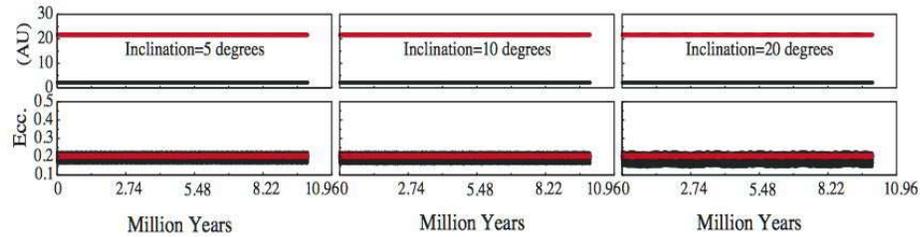}
\caption{Graphs of the semimajor axes (top) and eccentricities (bottom) 
of the giant planet (black) and binary (red) of $\gamma$ Caphei. 
The initial eccentricity of the binary at the beginning of numerical 
integration and the value of its mass-ratio were 
equal to 0.20 \citep{Hagh06}.}
\label{fig:7}
\end{figure}
\noindent
21.5 AU. Integrations also indicated that the binary-planetary
system of $\gamma$ Cephei becomes unstable in less than a few thousand years
when the initial value of the binary eccentricity exceeds 0.5. 

Interesting results were obtained when the $\gamma$ Cephei system was 
integrated for different values of $i_p$.
The results indicated that for the above-mentioned range of orbital
eccentricity, the planet maintains its orbit for all values of
inclination less than 40$^\circ$.  
Figure 9.7 shows the semimajor axes and orbital eccentricities
of the system for ${e_b}=0.2$ and for  $i_p$=5$^\circ$, 10$^\circ$,
and 20$^\circ$. 
For orbital inclinations larger than 40$^\circ$, the
system becomes unstable in a few thousand years.

\subsubsection {Kozai Resonance}

An interesting dynamical phenomenon that may occur
in an S-type binary, 
and has also been observed in the numerical simulations of
a few of these systems, is the Kozai resonance
\citep{Hagh03,Hagh05,Verrier06,Takeda06,Malmberg07}.
As demonstrated by \citet{Kozai62}, in a three-body system with
two massive objects and a small body (e.g., a binary-planetary
system), the exchange 
of angular momentum between the planet and the secondary star,
can cause the orbital eccentricity of the planet to reach high values 
at large inclinations. Averaging the equations of motion
of the system over mean anomalies, one can show that in this case,
the averaged system is integrable when the ratio of distances are 
large \citep[the Hill's approximation,][]{Kozai62}.
The Lagrange equations of motion in this case, indicate that,
to the first order of planet's eccentricity, the longitude of
the periastron of this object, $\omega_p$, librates
around a fix value. Figure 9.8 shows this for the giant
planet of $\gamma$ Cephei. As shown here, $\omega_p$ librates 
around 90$^\circ$ \citep{Hagh03,Hagh05}.  

In a Kozai resonance, the longitude of periastron and the 
orbital eccentricity of the small body $(e_p)$are related to its 
orbital inclination as \citep{Innanen97}
\begin{equation}
{\sin^2}{\omega_p}=\,0.4\,{\csc^2}{i_p},
\end{equation}
\noindent
and

\begin{figure}
\centering
\vskip 0.8in
\includegraphics[height=9cm]{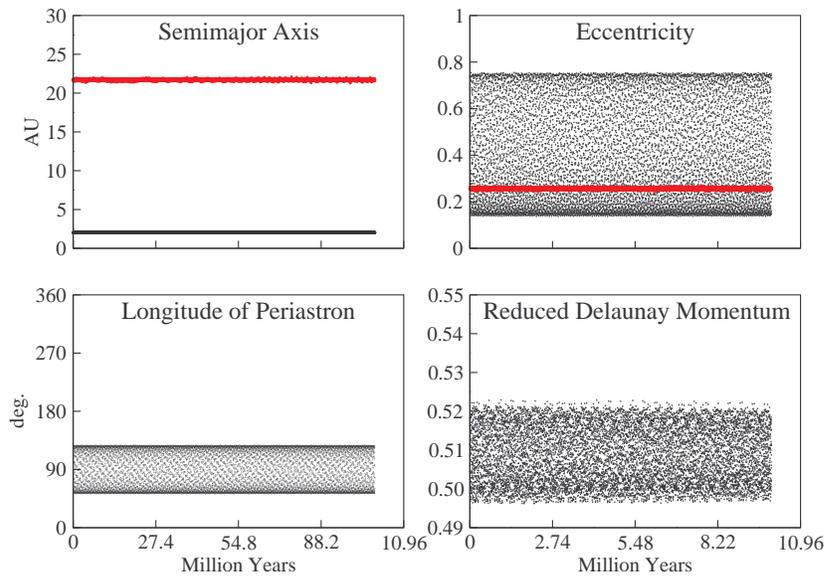}
\vskip -0.65in
\caption{Graphs of the semimajor axis and eccentricity of the giant planet
(black) and binary (red) of $\gamma$ Cephei (top) and its
longitude of periastron and reduced Delaunay momentum (bottom)
in a Kozai resonance 
\citep{Hagh03,Hagh05}. As expected, the longitude of the periastron of the
giant planet oscillates around $90^\circ$ and its reduced Delaunay
momentum is constant.}
\label{fig:8}
\end{figure}

\begin{equation}
{(e_p^2)_{\rm max}}={1\over 6}\,\Bigl[1-5\cos (2{i_p})\Bigr].
\end{equation}
\vskip 10pt
\noindent
From equation (9.3), one can show that the Kozai resonance may occur
if the orbital inclination of the small body is
larger than 39.23$^\circ$. For instance, as shown by \citet{Hagh03,Hagh05}, 
in the system of $\gamma$ Cephei, Kozai resonance occurs 
at ${i_p}={60^\circ}$.
For the minimum value of ${i_p}$, the maximum value
of the planet's orbital eccentricity, as given by equation (9.4), 
is equal to 0.764. Figure 9.8 also shows that
$e_p$ stays below this limiting value at all times.

As shown by \citet{Kozai62} and \citet{Innanen97}, 
in a Kozai resonance, the disturbing function
of the system, averaged over the mean anomalies, is independent of the
longitudes of ascending nodes of the small object (the planet) 
and the perturbing body (the stellar companion). 
As a result, the quantity ${\sqrt{{a}(1-{e^2})}}\,\cos {i}$
(shown as the ``Reduced Delaunay Momentum'' in figure 9.8)
becomes a constant of motion. Since the eccentricity and inclination 
of the planet vary with time, the

\begin{figure}
\centering
\vskip 0.2in
\includegraphics[height=4cm]{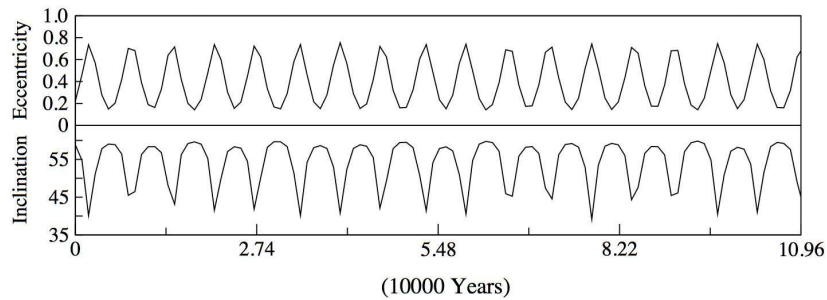}
\caption{Graphs of the eccentricity and inclination of the giant
planet of $\gamma$ Cephei in a Kozai Resonance \citep{Hagh03,Hagh05}. As
expected, these quantities have similar periodicity and are 
$180^\circ$ out of phase.}
\label{fig:9}
\end{figure}

\noindent
fact that the quantity above is a constant of motion implies that the
time-variations of these two quantities have similar periods
and, at the same time, they vary
in such a way that when $i_p$ reaches its maximum, $e_p$ reaches
its minimum and vice versa. Figure 9.9 shows this clearly.

\subsection {Stability of P-type Orbits}

Numerical simulations have also been carried out for the stability of 
P-type orbits in binary-planetary systems 
\citep{Ziglin75,Szebehely81,Dvorak84,Dvorak86,Dvorak89,Kubala93,
Holman99,Broucke01,Pilat03,Musielak05}. Similar to S-type orbits,
in order for a P-type planet to be stable, it has to be at a safe 
distance from the two stars so that it would be immune from their 
perturbative effects. That is, planets at large distances from the
center of mass of a binary will have a better chance of begin stable. 
This distance, however, cannot be too large because 
at very large distances,
other astronomical effects, such as galactic perturbation,
and perturbations due to passing stars, can render the orbit
of a planet unstable. 

To determine the critical value of the semimajor 
axis of a P-type planet in
a stable orbit, preliminary attempts were made by \citet{Dvorak84}, who
numerically integrated the orbit of a circumbinary planet
in a circular orbit around an eccentric binary system and
showed that planets at distances 2-3 times the separation
of the binary have stable orbits.
Subsequent studies by \citet{Dvorak86}, \citet{Dvorak89},
and \citet{Holman99} complemented Dvorak's results of
1984 and showed that the orbit of a P-type
planet will be stable as long as the semimajor axis of the planet
stays larger than the critical value given by (figure 9.10)

\begin{figure}
\centering
\vskip 0.4in
\includegraphics[height=8cm]{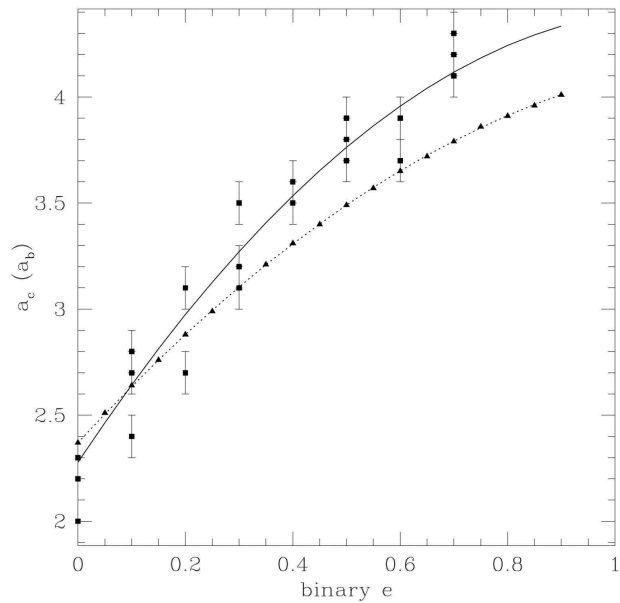}
\caption{Critical semimajor axis as a function of the binary eccentricity in 
a P-type system \citep{Holman99}. The squares correspond to the result of 
stability simulations by \citet{Holman99} and the triangles represent those
of \citet{Dvorak89}. The solid line corresponds to equation (9.4). As indicated
by \citet{Holman99}, the figure shows that at outer regions, the stability 
of the system fades away.}
\label{fig:10}
\end{figure}
\vskip 1pt
\vskip 0.5in
\begin{eqnarray}
&\!\!\!\!\!\!\!\!\!\!\!\!\!\!\!\!\!\!\!\!\!\!\!\!\!\!
{a_c}/{a_b}=(1.60 \pm 0.04) + (5.10 \pm 0.05){e_b}
+ (4.12 \pm 0.09)\mu \nonumber \\
&\qquad\qquad
+ (-2.22 \pm 0.11){e_b^2} + (-4.27 \pm 0.17){e_b}\mu 
+ (-5.09 \pm 0.11){\mu^2} \nonumber \\
&\!\!\!\!\!\!\!\!\!\!\!\!\!\!\!\!\!\!\!\!\!\!
\!\!\!\!\!\!\!\!\!\!\!\!\!\!\!\!\!\!\!\!\!\!\!\!\!\!\!\!\!
\!\!\!\!\!\!\!\!\!\!\!\!\!\!\!\!\!\!\!\!\!\!\!\!
+ (4.61 \pm 0.36){e_b^2}{\mu^2}\>.
\end{eqnarray}
\vskip 20pt

\noindent
Similar to equation (9.1),
equation (9.4) represents a transitional region with a lower
boundary below which the orbit of a P-type planet will be certainly
unstable, and an upper boundary beyond which the orbit of the
planet will be stable. The {\it mixed zone} 
between these two 
boundaries represents a region where a planet,
depending on its orbital parameters, and the
orbital parameters and the mass-ratio of the binary,
may or may not be stable. Recently, by applying the stability criteria 
of equation (9.4) to their observational results, 
\citet{Trilling07} have confirmed the presence of stable
dust band, possibly resulted from collision of planetesimal, around close binary
star systems.

A dynamically interesting feature of the stable region around
the stars of a binary is the appearance of islands
of instability. As shown by \citet{Holman99}
islands of instability may develop beyond the inner boundary of the 
{\it mixed zone}, which correspond to the locations of $(n:1)$
mean-motion resonances. The appearance of these unstable regions
have been reported by several authors 
under various circumstances \citep{Henon70,Dvorak84,Dvorak88a,Dvorak89}.
Extensive numerical simulations would be necessary to determine
whether the overlapping of these resonances would result
in stable P-type binary-planetary orbits.

\section {Planet Formation in Binaries}

Despite a wealth of articles on planets in binary star 
systems, the process of the formation of these objects
is still poorly understood. 
The current theories of planet formation focus only on the 
formation of planets in a circumstellar disk around a single star, 
and their extensions to binary environments are limited to
either the Sun-Jupiter system, where the focus is
on the effect of Jupiter on the formation of inner planets 
of our solar system
\citep{Hep74,Hep78,Drob78,Diakov80,Whitmire98,Kortenkamp01}, 
or binaries resembling some of
extrasolar planets, in which the secondary star has a mass in the 
brown dwarf regime \citep{Whitmire98}. 
Although attempts have been made to extend such studies to binaries with
comparable-mass stellar components
\citep{Marzari00,Nelson00,Barbieri02,Quintana02,Lissauer04}, 
the extent of the applicability of the results of these studies
has been only to hypothetical cases since, until recently,
there had been no observational evidence on the existence of
such binary-planetary systems.

In general, it is believed that planet formation
proceeds through the following four stages (figure 9.11):

1) coagulation of dust particles and their growth to centimeter-sized
objects,

2) collisional growth of centimeter-sized particles to 
kilometer-sized bodies (planetesimals),

3) formation of Moon- to Mars-sized protoplanets (also known as 
planetary embryos) through the collision and coalescence of 
planetesimals, and

4) collisional growth of planetary embryos to 
terrestrial-sized objects. 
\hfill

\noindent
The latter is a slow process
that may take a few hundred million years. During the first few 
million years of this process, at larger distances from the star,
planetesimals and planetary embryos may form planetary cores several times
more massive than Earth, and may proceed to form giant planets.

In a binary star
system with a moderate to small separation, the secondary
star will have significant effects on the efficiency of each
of these processes. As shown by Boss (2006), a binary companion
can alter the structure of a planet-forming nebula, and create
regions where the densities of the gas and dust are locally enhanced
(figure 9.12). Also, as shown by  \citet{Artymowicz94}, and
\begin{figure}
\centering
\vskip 0.1in
\includegraphics[height=7.5cm]{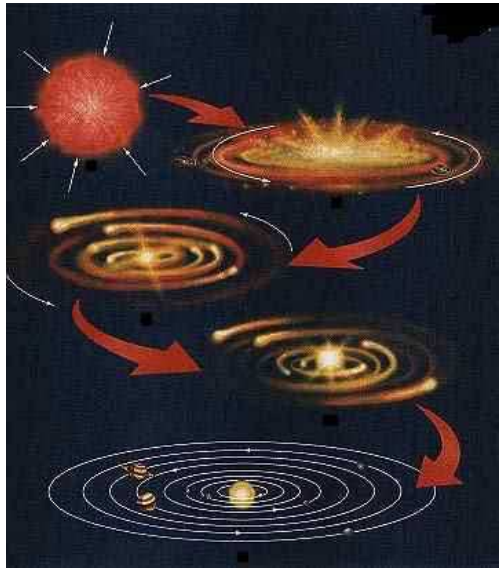}
\caption{The four stages of planet formation.}
\label{fig:11}
\end{figure}
\begin{figure}
\centering
\vskip 0.2in
\includegraphics[height=7cm]{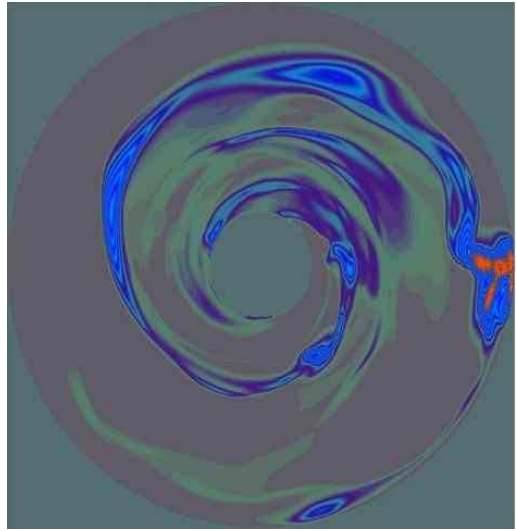}
\caption{Structure of a circumprimary disk in a double star system.
The masses of the primary and secondary stars are 1 and 0.09 solar-masses,
respectively. The secondary star is at 50 AU at the top of the figure,
and has an eccentricity of 0.5. The figure shows an area of 20 AU around
the primary. The structures inside the disk have appeared after
239 years from the beginning of the simulations. 
The orange structure on the right edge of the graph is
an artifact of numerical simulations \citep{Boss06}.}
\label{fig:12}
\end{figure}
\begin{figure}
\centering
\vskip 0.7in
\includegraphics[height=3cm]{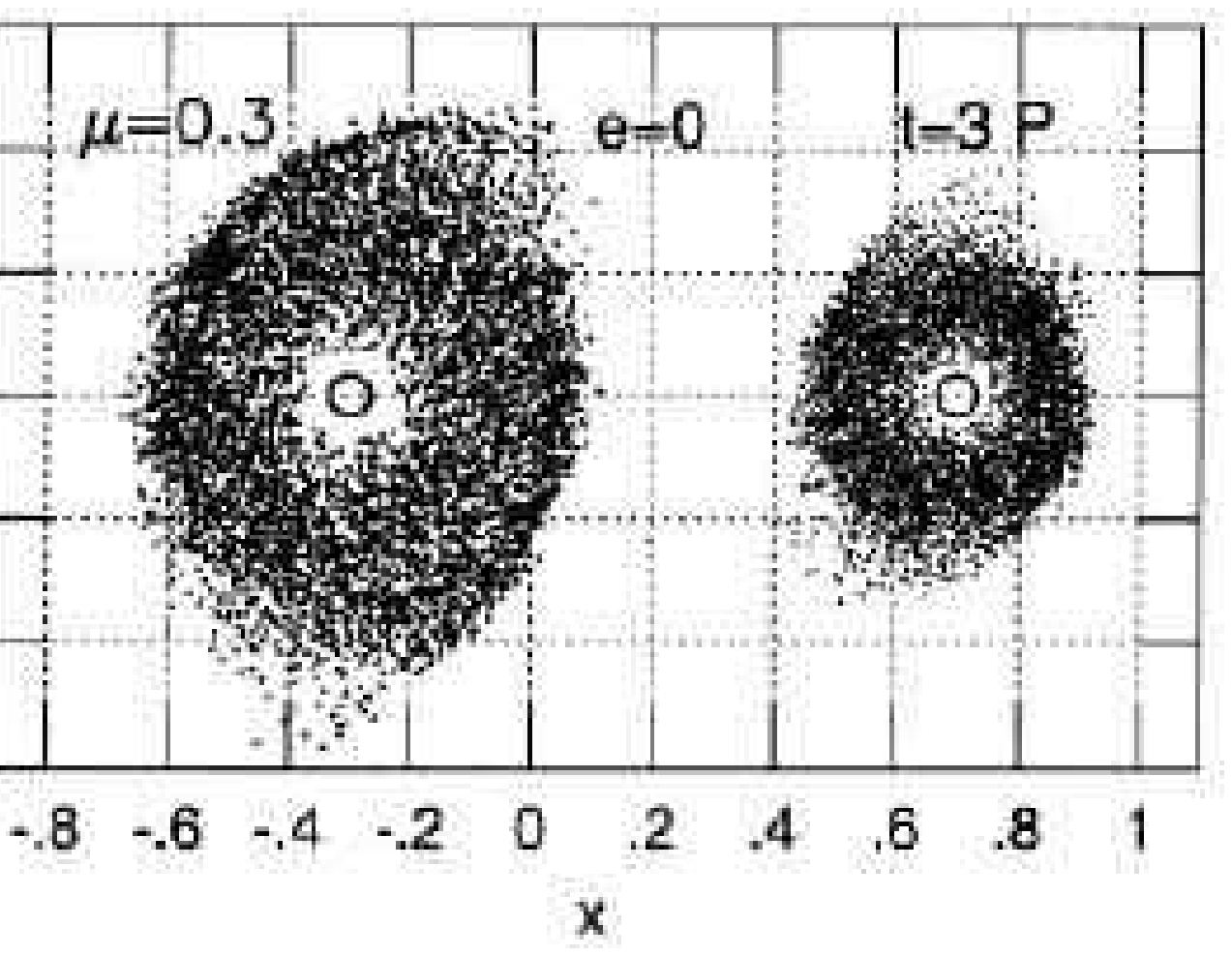}
\includegraphics[height=3cm]{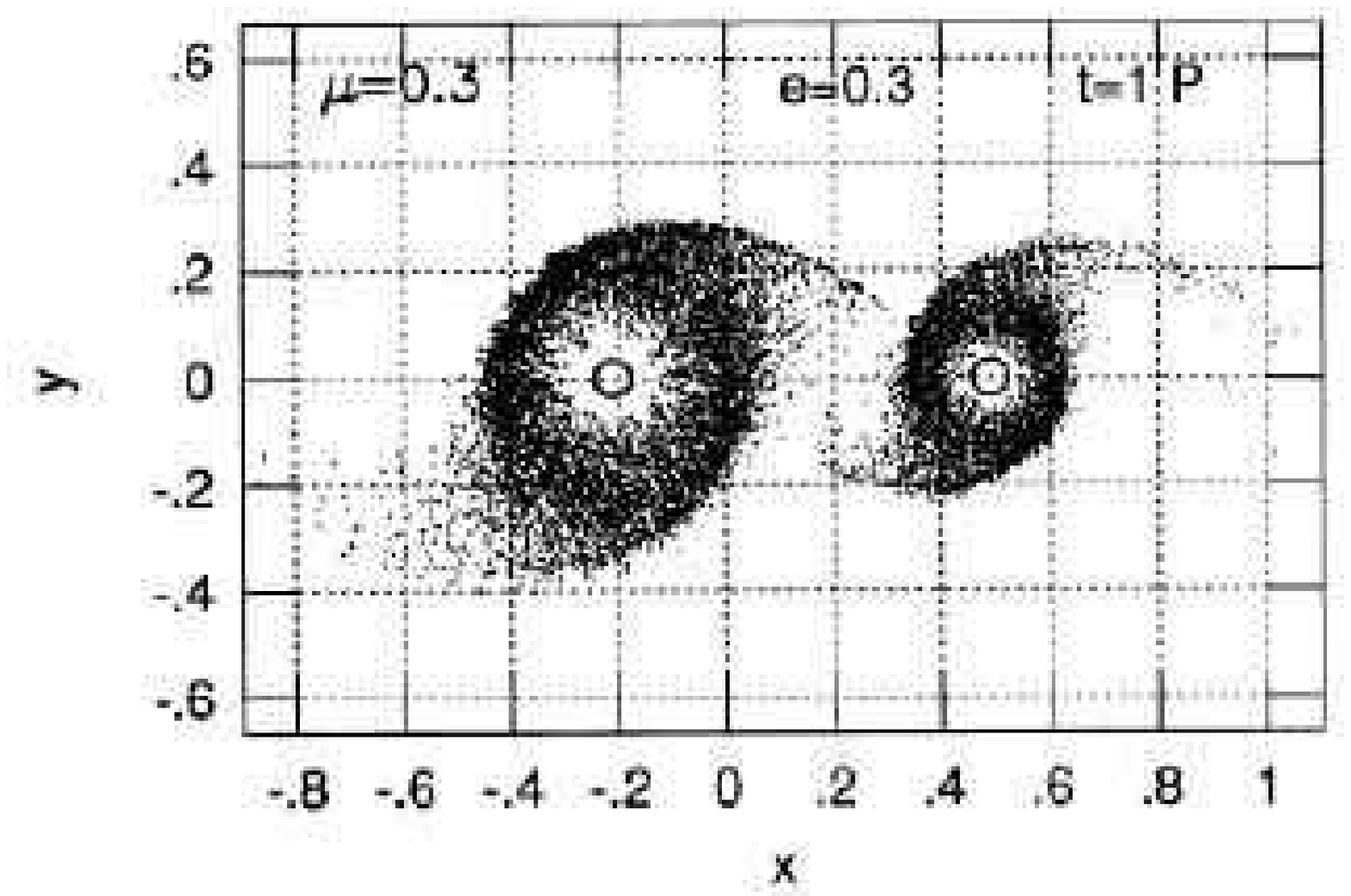}
\vskip 20pt
\includegraphics[height=11cm]{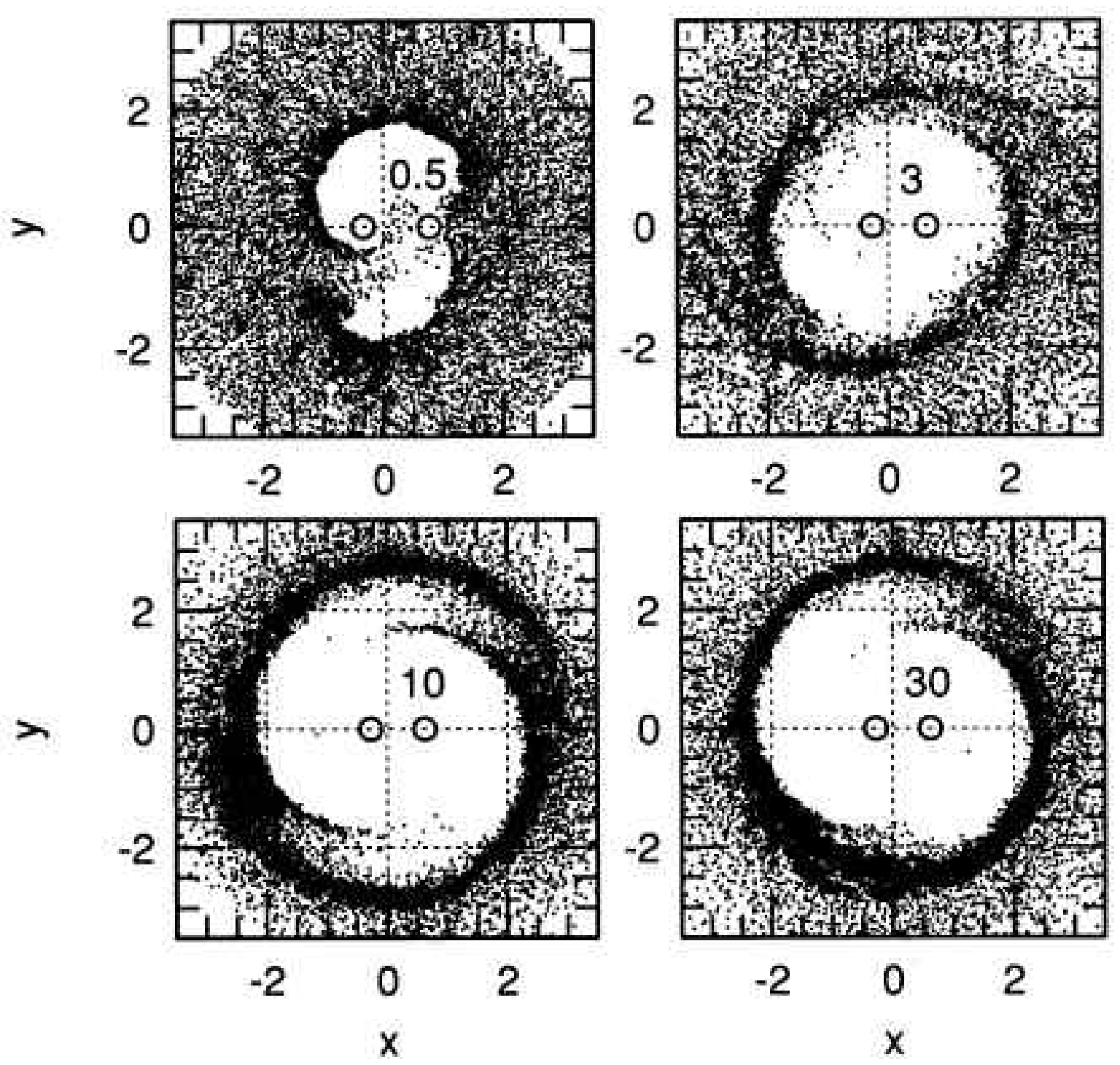}
\caption{Disk Truncation in and around binary systems
\citep{Artymowicz94}. The top graphs show circumstellar disk in a binary
with a mass-ratio of 0.3. Note the disk truncation when the eccentricity
of the binary is increased from 0 to 0.3. The bottom graphs show
similar effect in a circumbinary disk. The mass-ratio is 0.3 and the
binary eccentricity is 0.1. The numbers inside each graph represent 
the time in units of the binary period. The axes are in units
of the binary semimajor axis.}
\label{fig:13}
\end{figure}
\begin{figure}
\centering
\includegraphics[height=10cm]{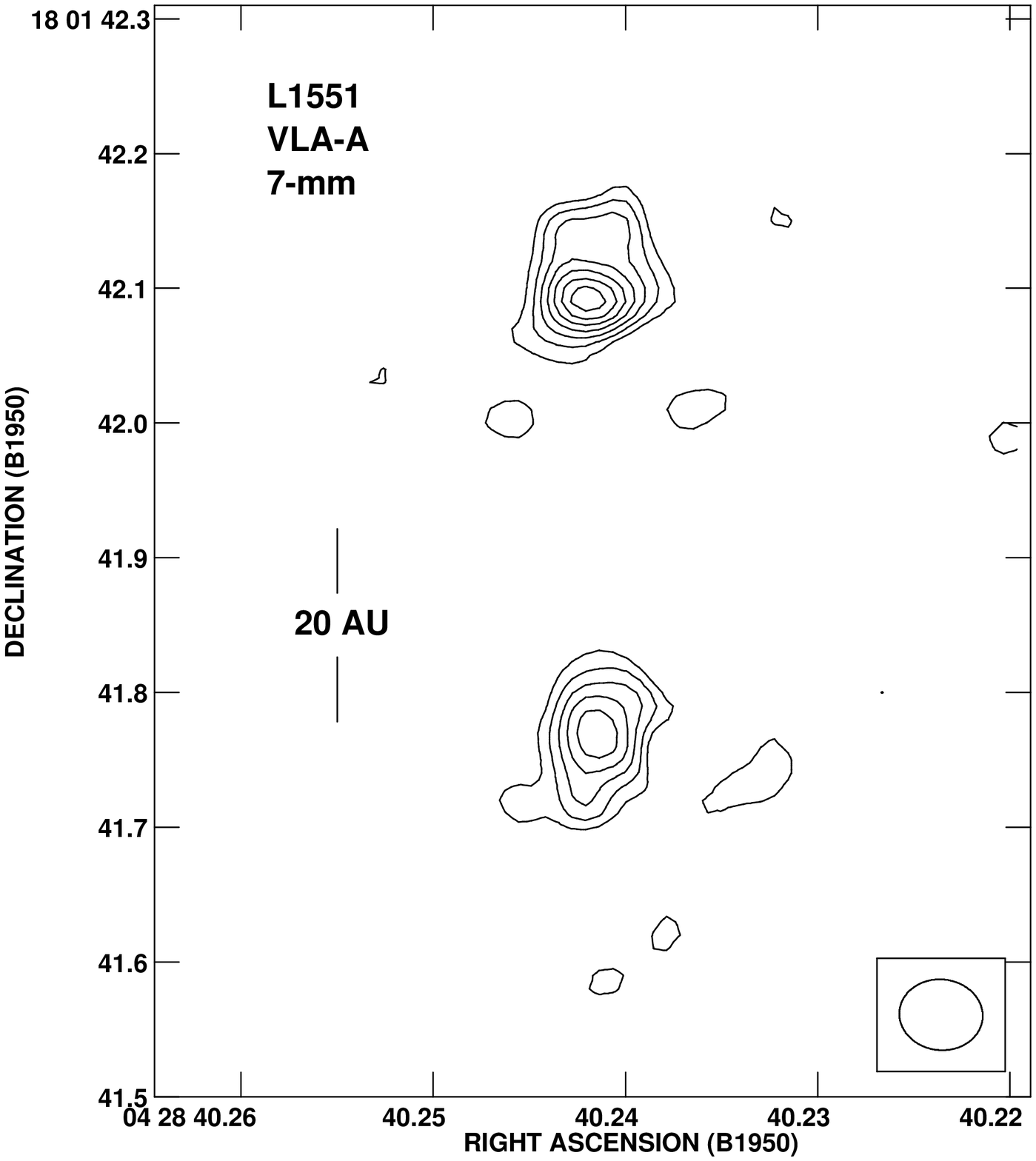}
\vskip -10pt
\caption{Interferometric observation of the binary system L1551
\citep{Rodriguez98}. Two compact sources are evident in the map.
The separation of the binary is 45 AU and the disk around each 
core extends to approximately 10 AU.}
\label{fig:14}
\end{figure}
\noindent
\citet{Pichardo05},
a stellar component on an eccentric orbit can truncate the
circumprimary disk of embryos to smaller radii and remove 
material that may be used in the formation of terrestrial planets
(figure 9.13). As a result, 
it used to be believed that circumstellar
disks around the stars of a binary may not be massive enough to
form planets. 
However, observations by \citet{Math94},
\citet{Akeson98}, \citet{Rodriguez98}, and \citet{Math00}
have indicated that potentially planet-forming circumstellar disks
can indeed exist around the stars of a binary system, implying that 
planet formation in binaries  may be as common as around single 
stars (figure 9.14). Among these
circumstaller disks, the two well-separated disks of the system
L1551 retain equivalent of approximately 0.03 to 0.06 solar-masses 
of their original circumstellar materials in a region with an outer
radius of $\sim$10 AU \citep[figure 9.14, ][]{Rodriguez98}. 
The masses of these disks are
comparable to the minimum solar-mass model of the primordial nebula
of our solar system \citep{Stu77,Hayashi81}, implying that, planet
formation in dual-star systems can begin and continue in the same
fashion as around our Sun.

Despite the observational evidence in support of the existence 
of planet-forming environments in moderately close 
binary star systems, the perturbative effect of the 

\begin{figure}
\centering
\vskip -0.15in
\includegraphics[height=6cm]{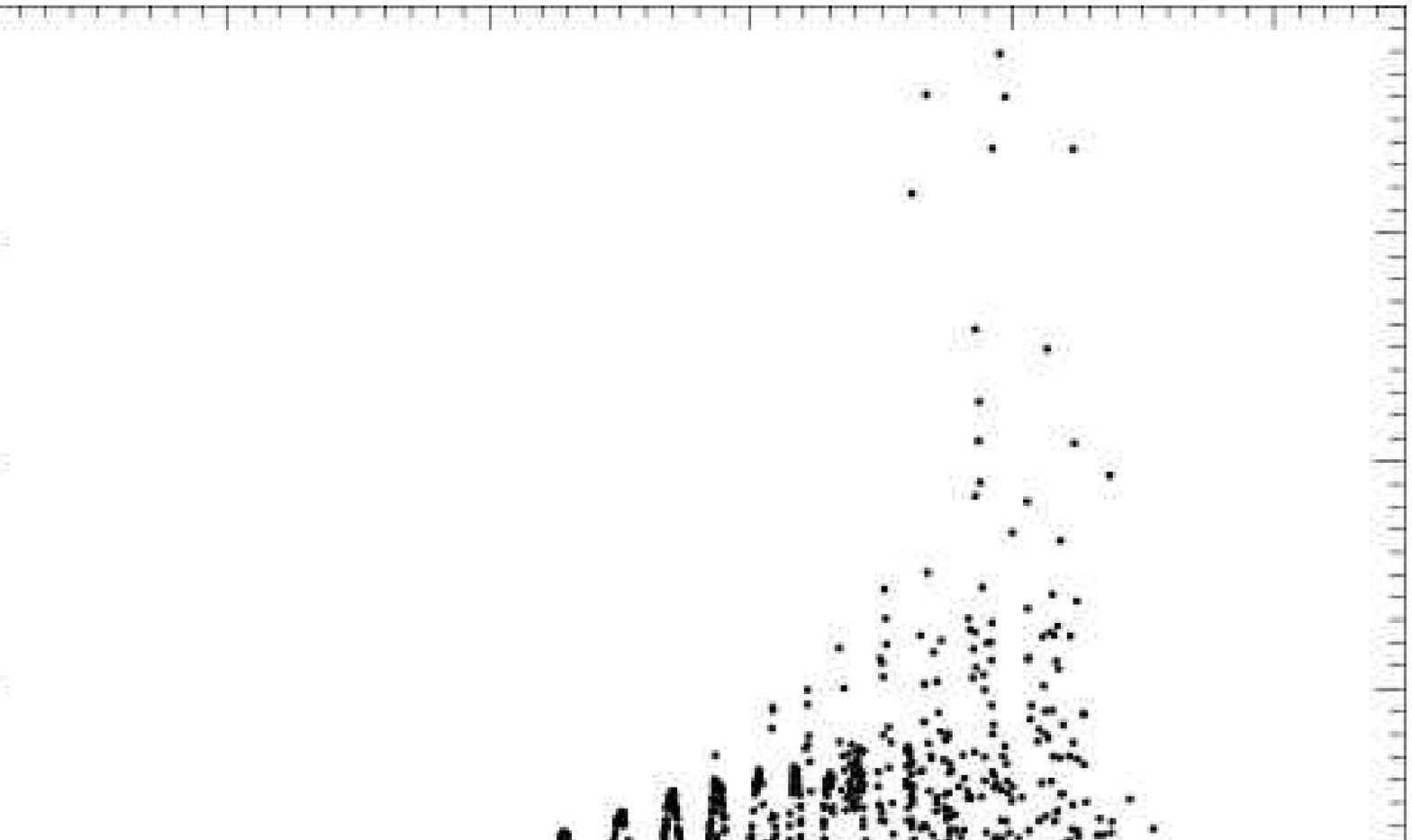}
\vskip 10pt
\includegraphics[height=6cm]{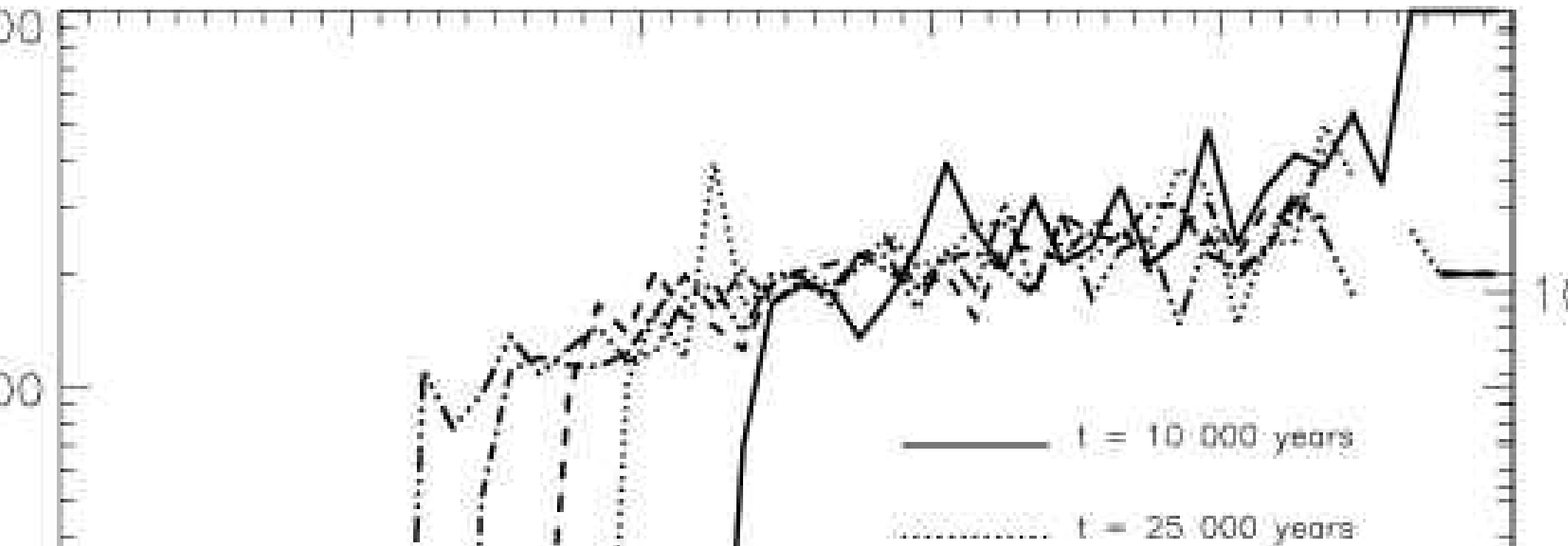}
\vskip 10pt
\vskip 1.8in
\caption{Graphs of the evolution of eccentricity (top)
and encounter velocities (bottom)
for planetesimals at the region between 0.3 and 5 AU from the primary of
$\gamma$ Cephei \citep{Thebault04}. The planetesimals disk in the
bottom simulation was initially at its truncated radius of 4 AU.
As shown here, the perturbative effect
of the secondary star increases the eccentricities and relative velocities
of these objects.}
\label{fig:15}
\end{figure}
\begin{figure}
\centering
\vskip 0.15in
\includegraphics[height=7cm]{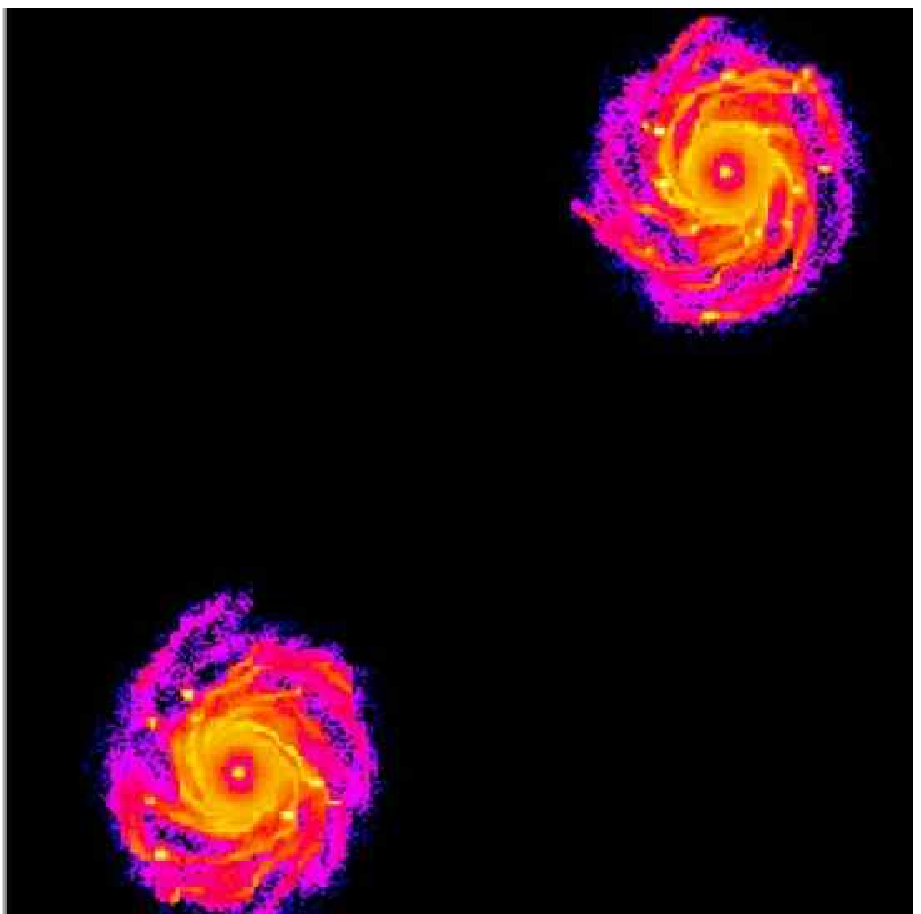}
\caption{Giant planet formation via disk instability mechanism
in a binary system. The separation of the binary is 120 AU and
it was initially on a circular orbit. The mass of each
disk is 0.1 solar-masses. The snap shot was taken 160 years
after the start of the simulations. Figure Courtesy of L. Mayer,
A. Boss and A. Nelson .}
\label{fig:16}
\end{figure}

\noindent
binary companion
may not always favor planet formation. For instance,
as shown by \citet{Nelson00}, giant planet formation cannot
proceed through the disk instability mechanism \citep{Boss00}
around the primary of
a binary star system with separation of $\sim 50$ AU. Also, when forming
planetary embryos, as shown by \citet{Hep78}, 
\citet{Whitmire98}, and \citet{Thebault04},
the perturbation of
the secondary star
may increase the relative velocities of planetesimals and
cause their collisions to result in breakage and fragmentation (figure 9.15). 
Results of the studies by these authors suggest that 
planetesimal accretion will be efficient only in binaries
with large separation [50 AU as indicated by \citet{Hep78}, 
26 AU as shown by \citet{Whitmire98}, and 100 AU as reported by
\citet{Mayer05}]. Finally, in a binary
star system, the stellar companion may 
create unstable regions where the building blocks of planets
will not maintain their orbits and, as a result, planet formation
will be inhibited \citep{Whitmire98}. 

Interestingly, despite all these difficulties, numerical simulations have
shown that it may indeed be possible to form giant and/or terrestrial planets
in and around a dual-star system. Recent simulations
by \citet{Boss06}, and \citet{Mayer07} 
indicate that Jupiter-like planets can form around
the primary of a binary star system via gravitational instability 
in a marginally unstable circumprimary disk (figure 9.16).
On the other hand,
as shown by \citet{Thebault04}, the core 
accretion mechanism may also
be able to form giant planets around the primary of a binary star.
However, as the results of their simulations for planet formation in
\begin{figure}
\centering
\vskip 0.2in
\includegraphics[height=8cm]{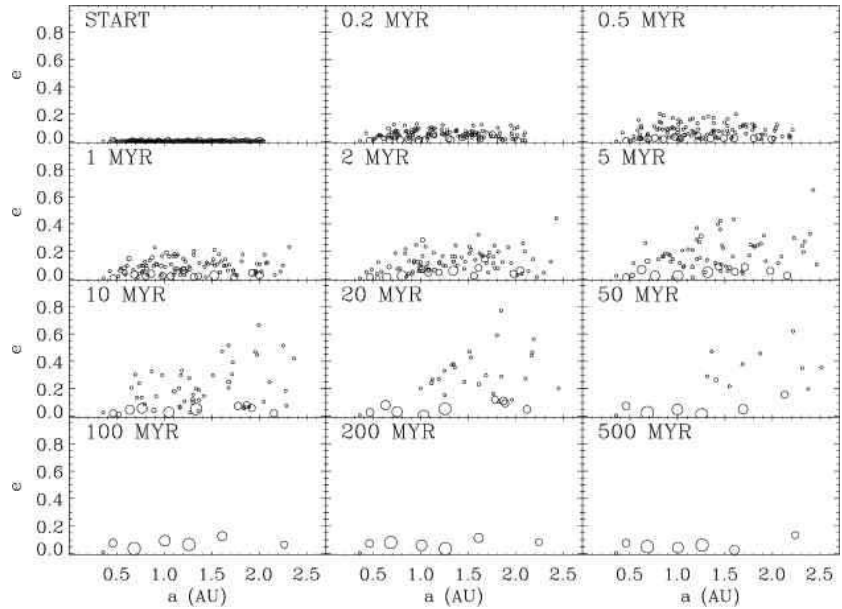}
\caption{Terrestrial planet formation around a close binary system
\citep{Quintana06}. The binary is circular and its separation is
0.05 AU. Each star of the binary has a mass of 0.5 solar-masses.
A Jupiter-like planet has also been included in the simulation. 
The circles represent planetary embryos and planetesimals with radii
that are proportional to their physical sizes. As shown here, the
perturbative effect of the outer giant planet excite the orbits
of the bodies at the outer edge of the disk and causes radial mixing
as well as truncation. Within the first 100 Myr, several terrestrial-class
objects are formed around the binary system.}
\label{fig:17}
\end{figure}

\noindent
the $\gamma$ Cephei system indicate, the semimajor axis of the final gas-giant
planet may be smaller than its observed value. 

In regard to the formation of terrestrial planets in binary systems, 
in a series of
articles, Quintana and her colleagues integrated the orbits of
a few hundred Moon- to Mars-sized objects and showed that 
terrestrial-class objects can form in and around binaries
\citep{Quintana02,Lissauer04,Quintana06,Quintana07}.
Figure 9.17 shows the results of some of their
simulations. As shown here, depending on the mass-ratio of the
binary and the initial values of its orbital parameters, 
in a few hundred million years, terrestrial planets can form 
around a close (0.01 to 0.1 AU) binary star system.

Quintana and colleagues also studied terrestrial planet formation
in binaries with larger separations \citep{Quintana07}. Figure 9.18
shows the results of their simulations for a binary with a separation
of 20 AU. Similar to figure 9.17, 
terrestrial-type objects are formed
around the primary of the binary in a few hundred million years.
Statistical analysis of their results, as shown in figures 9.19, 

\begin{figure}
\centering
\vskip 0.1in
\includegraphics[height=11cm]{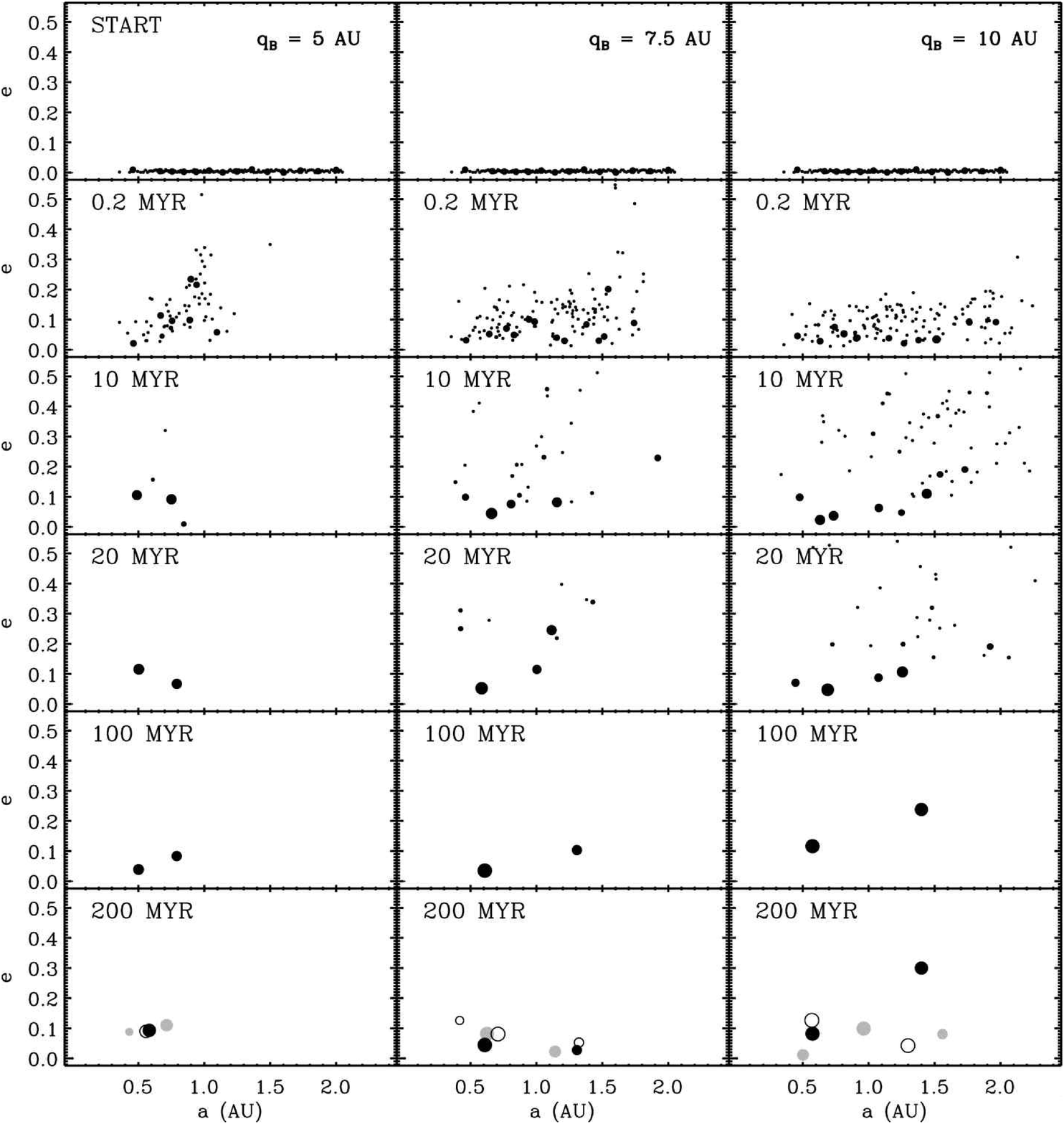}
\caption{Terrestrial planet formation around the primary of a binary 
star system \citep{Quintana07}. The stars of the binary are 
0.5 solar-mass with semimajor axis of 20 AU. The eccentricity of 
the binary is 0.75 (left column), 0.625 (middle column), 
and 0.5 (right column). As shown here, in each simulations,
two terrestrial-type objects are formed after 100 Myr. The last
row shows the results of additional simulations of the same
systems, with final results showing in black, gray and white,
corresponding to different runs. The differences in the final
assembly of the planetary system of each simulation are results
of the stochasticity of this type of numerical integrations.}
\label{fig:18}
\end{figure}
\vskip 10pt
\noindent
9.20, and 9.21 indicate that, as expected
in binaries with larger perihelia, where disk truncation has been smaller 
and more planet-forming material is available, terrestrial planet
formation is efficient and the number of final terrestrial
planets is large. The results of simulations by \citet{Quintana07}
also indicate that in a binary with a periastron
distance larger than 10 AU, terrestrial planet formation can
proceed efficiently in a region within 2 AU of the primary
star. In binaries with periastra smaller than 5 AU, this
region may be limited to inside 1 AU (figure 9.20).

\begin{figure}
\centering
\vskip 0.1in
\includegraphics[height=6.55cm]{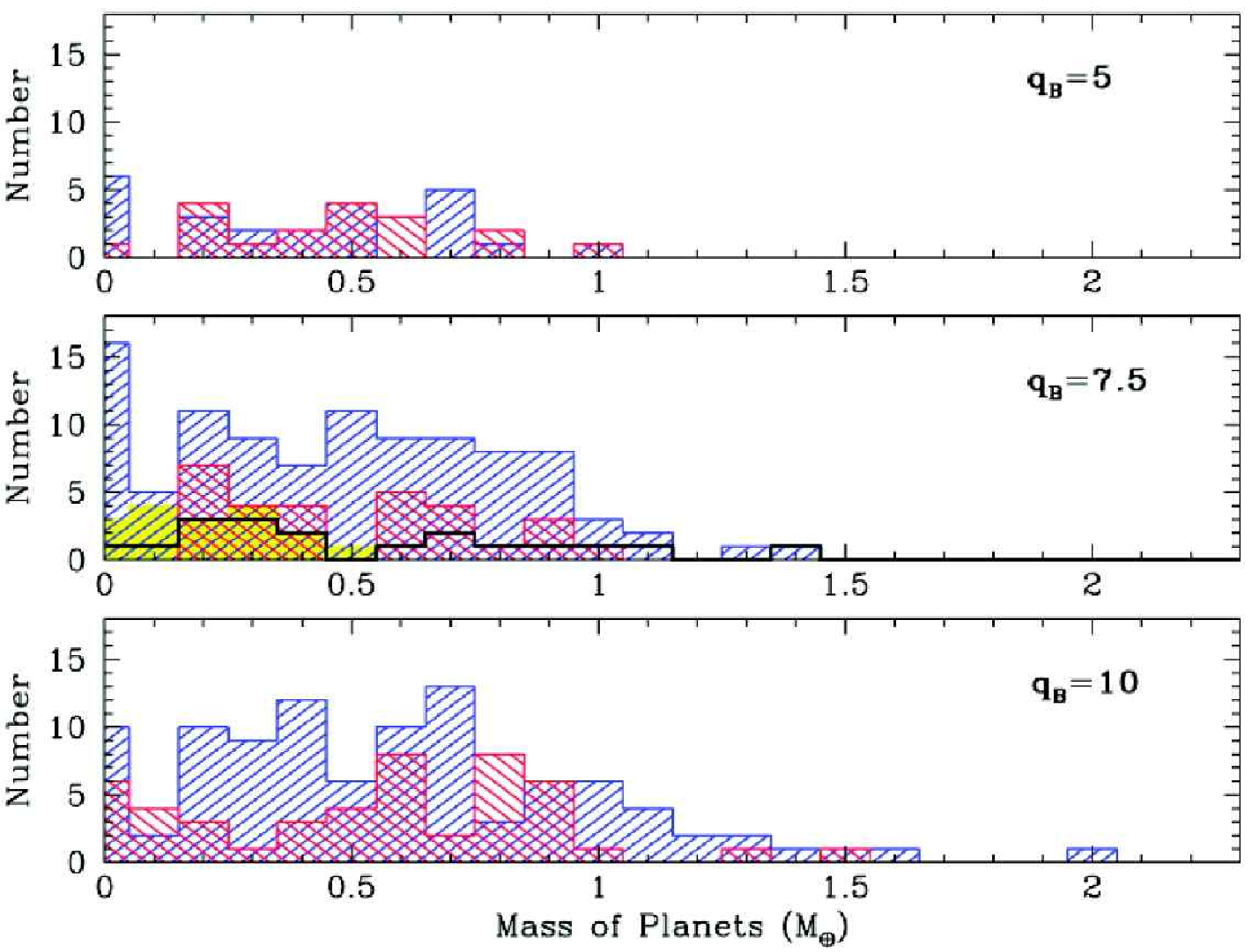}
\caption{Graphs of the final masses of the terrestrial planets 
formed in systems of figure 9.18. The simulations have been
run for three different masses of the binary stars. The red 
corresponds to simulations in a binary with 0.5 solar-masses stars,
the blue represents results in a binary with 1 solar-mass stars,
yellow is for a binary with a 1 solar-mass primary and a 0.5 solar-masses 
secondary, and 
black represents a binary with a 0.5 solar-masses primary and a 1 solar-mass 
secondary. These results show that despite the disk truncation
in binaries with smaller perihelia, the average masses of the final
planets are not significantly altered.
\citep{Quintana07}.}
\label{fig:19}
\end{figure}
\begin{figure}
\centering
\includegraphics[height=6.55cm]{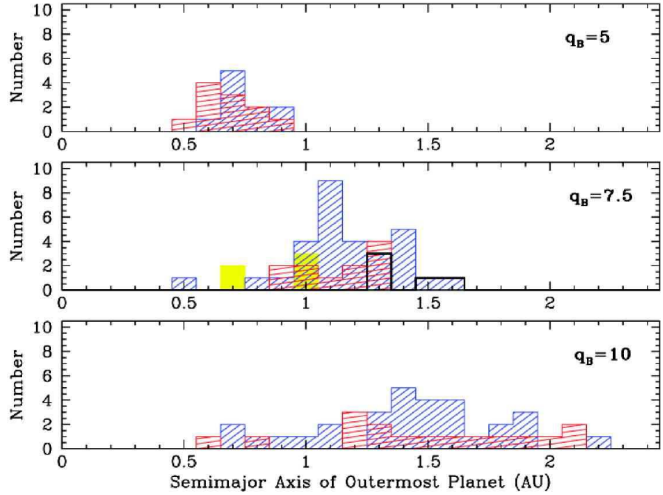}
\caption{Graphs of the semimajor axis of the outermost 
planet of the simulations of figure 9.18. 
A shown here, while the outer edge of the disk is affected by the
presence of the binary companion, the inner portion of the disk,
where terrestrial planets are formed,
stays unaffected by this object \citep{Quintana07}.}  
\label{fig:20}
\end{figure}

\begin{figure}
\centering
\includegraphics[height=6.55cm]{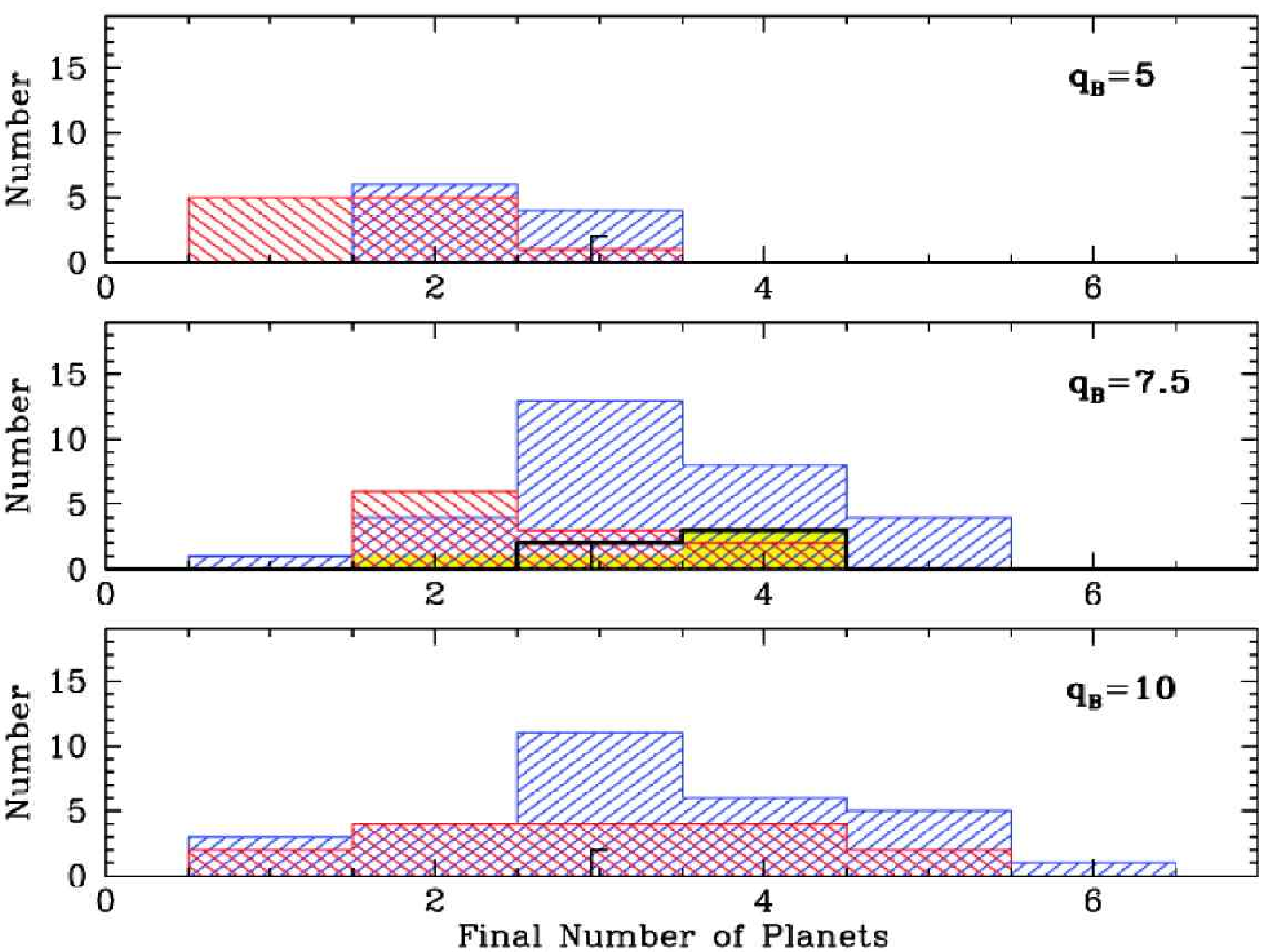}
\caption{Number of final terrestrial-type planets formed 
in the binaries of figure 9.18. As expected, for a given binary
mass-ratio, the number of terrestrial planets increases  
in systems with larger perihelia. This number also increases
when the mass of the binary companion is smaller.
\citep{Quintana07}.}
\label{fig:21}
\end{figure}

Despite the destructive role of the binary companion in increasing
the relative velocities of planetesimals, which causes their collisions
to result in erosion, this efficiency of terrestrial planet formation 
in binary systems may be attribute to the fact that
the effect of the binary companion on increasing
the relative velocities of planetesimals can be counterbalanced
by dissipative forces such as gas-drag and dynamical friction
\citep{Marzari97,Marzari00}.
The combination of the drag force of the gas and the gravitational
force of the secondary star may result in the alignment of the periastra
of planetesimals and increases the efficiency of their accretion
by reducing their relative velocities. This is a process that is 
more effective when the sizes
of the two colliding planetesimals are comparable and small. For colliding
bodies with different sizes, depending on the size distribution of
small objects, and the radius of each individual planetesimal,
the process of the alignment of periastra may instead increase 
the relative velocities of the two objects, and cause their 
collisions to become eroding \citep[figure 9.22, also see][]{Thebault06}. 

Simulations of terrestrial
planet formation have also been extended to binaries with
larger separations (20-40 AU) that also host a giant planet
\citep{Hagh07}. As discussed in the next section,
by numerically integrating the orbits of the binary, its giant planet,
and a few hundred planetary embryos, these authors have shown
that it is possible to form Earth-like objects, with considerable
amount of water, in the habitable zone of the primary of a moderately
close binary-planetary system.

\begin{figure}
\centering
\vskip 0.1in
\includegraphics[height=7cm]{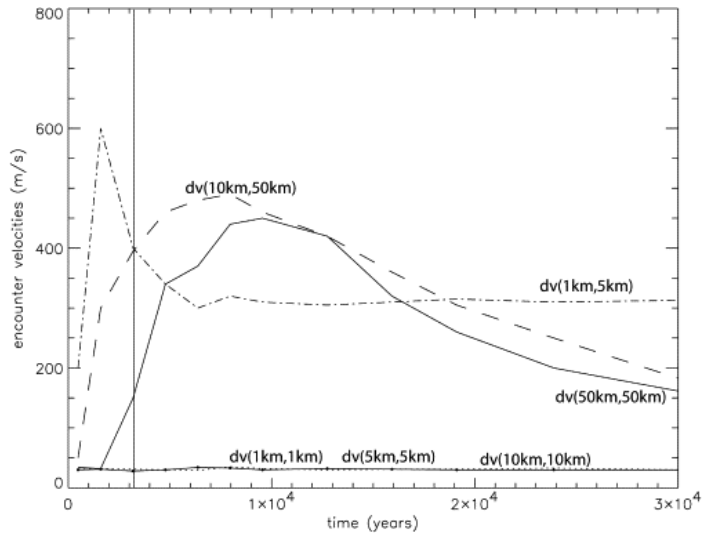}
\caption{Encounter velocities of planetesimals with different sizes
in a binary system with semimajor axis of 10 AU, eccentricity of 0.3,
and mass-ratio of 0.5 \citep{Thebault06}. The simulations include 
gas-drag. The vertical line shows the time of orbital crossing. 
As shown here, gas-drag lowers the encounter velocities of smaller 
equal-size planetesimals through the periastron alignment process.}   
\label{fig:22}
\end{figure}

\section {Habitability}

It is widely accepted that a planet capable of developing 
life (as we know it), has to be able to 
continuously maintain liquid water on its surface and in 
its atmosphere. 
The capability of a planet in retaining water
depends on its size and the processes involving its interior dynamics and 
atmospheric circulation. It also depends on its orbital 
parameters (i.e., its semimajor axis and
orbital eccentricity) and the brightness of the central star at the
location of the planet. While a dynamic interior and atmospheric
circulation are necessary for a habitable planet to develop CO$_2$ cycle and
generate greenhouse effect (which helps the planet to maintain
a uniform temperature), a long-term stable orbit,
at a right distance from the star, is essential to ensure
that the planet will receive the amount of radiation that
enables it to maintain liquid water on its surface. These
seemingly unrelated characteristics of a potential habitable planet,
have strong intrinsic correlations, and combined with the
luminosity of the star, determine the system's {\it habitable zone}.

The inner and outer boundaries of a habitable zone vary with
the star's luminosity and the planet's atmospheric circulation
models [see \citet{MnT03}, \citet{Jones05}, 
and \citet{Jones06} for a table of distances of the
inner and outer boundaries of the habitable zones of exoplanetary
systems]. A conservative estimate of the habitable zone of a star
can be made by assuming that its inner edge is located at
a distance closer than which
\begin{figure}
\centering
\vskip 0.1in
\includegraphics[height=9cm]{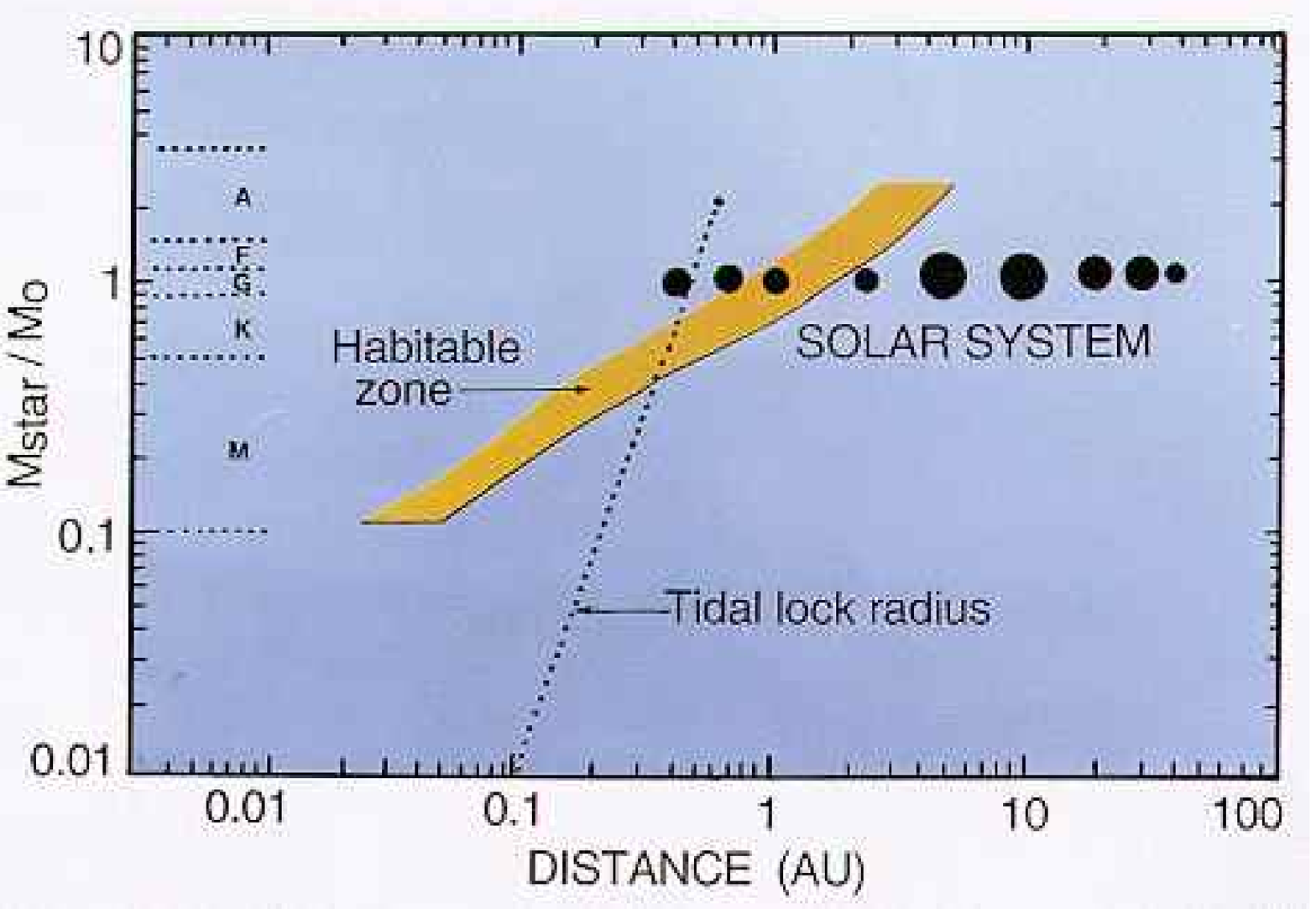}
\caption{Habitable zone
\citep{Kasting93}.}
\label{fig:23}
\end{figure}
\vskip 10pt
\noindent
water on the surface of the planet 
evaporates due to runaway greenhouse effect, and its outer edge 
is at a distance where, in the absence
of CO$_2$ clouds, runaway glaciation will freeze the water and creates
permanent ice on the surface of the planet. As shown by \citet{Kasting93}, 
such a definition of a habitable zone results in a habitable region
between 0.95 AU and 1.15 AU for the Sun 
(figure 9.23). This is a somewhat conservative
estimate of the Sun's habitable zone and as noted by \citet{Jones05}, 
the outer edge of this region may, in fact, be farther away
\citep{Forget97,Williams97,Mischna00}.

Since the notion of habitability is based on life on Earth,
it is possible to calculate the location of the boundaries of 
the habitable zone of a star by comparing its luminosity with 
that of the Sun. For a star with the surface temperature 
$T$ and radius $R$, the luminosity $L$ is given by
\begin{eqnarray}
{L}({R},{T})=4\pi\sigma{T^4}{R^2},
\end{eqnarray}
\noindent
where $\sigma$ is Boltzmann constant.
Using equation (9.5) and the fact that Earth 
is in the habitable zone of the Sun,
the radial distances of the inner and outer edges of the habitable 
zone of a star can be obtained from (Haghighipour 2006)
\begin{equation}
r\,=\,
{\Bigl({{T}\over {T_{\odot}}}\Bigr)^2}\,
{\Bigl({{R}\over {R_{\odot}}}\Bigr)}\,
{r_{\odot}}.
\end{equation}

\begin{figure}
\centering
\vskip 0.2in
\includegraphics[height=7cm]{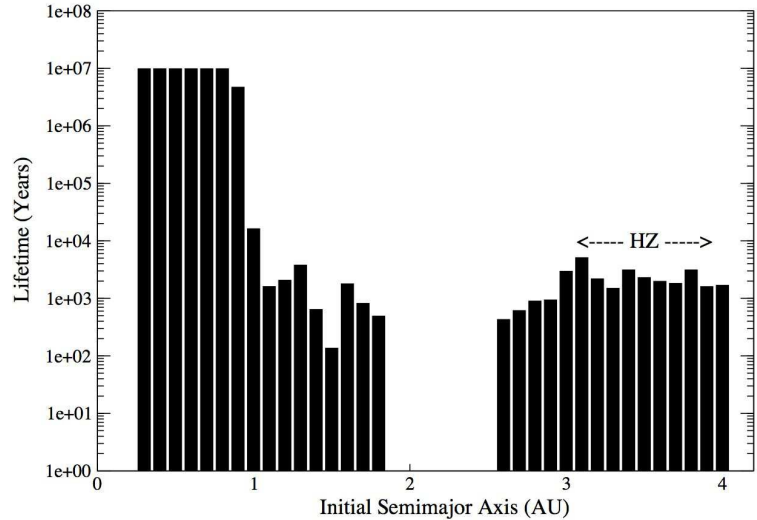}
\caption{Graph of the lifetime of an Earth-size object in
a circular orbit around the primary of $\gamma$ Cephei. The
habitable zone of the primary has been indicated by HZ. No 
planet was placed in the region between the aphelion and
perihelion distances of the giant planet of the system. As
shown here, only Earth-size planets close to the primary star
maintain their orbits for long times 
\citep{Hagh06}.}
\label{fig:24}
\end{figure}
\noindent
The quantities 
$T_{\odot}$ and $R_{\odot}$ in equation (9.6) are the surface
temperature and radius of the Sun, respectively, and $r_{\odot}$
represent the radial distance of Earth from the Sun.
Equation (9.6) implies that
a habitable zone can be defined as a region around a star where  
an Earth-like planet can receive the same amount of radiation as Earth
receives from the Sun, so that it can develop and 
maintain similar habitable 
conditions as those on Earth. 

As mentioned above, the orbit of a potential habitable planet 
in the habitable zone of a star has to be stable over long durations
of time. As shown in section 9.2, the stability of the orbit of a planet
in a binary system is strongly affected by the orbital motion
of the binary companion. In binary systems where the primary 
hosts other planetary bodies (e.g., giant planets), 
the dynamics of a habitable planet will also be affected by the
gravitational perturbations of these objects. 
It is therefore important to determine under what conditions a
terrestrial-class object will have a long-term stable orbit
in the habitable zone of a binary system, prior to
construction a theory for the formation of Earth-like planets 
in such environments.

Since a terrestrial-class planet is
approximately two orders of magnitude less massive than
a Jovian-type object, it will not have significant effect
on the motion of the stars and the giant planets of a binary system.
Therefore, as explained in sections 9.2.1 and 9.2.2, 
if a binary system does not contain a Jupiter-like
planet, any dynamical criterion that is obtained for the stability
or instability of a general planetary body,
\begin{figure}
\centering
\vskip -0.3in
\includegraphics[height=6.4cm]{Fig25a.eps}
\vskip 10pt
\includegraphics[height=6.4cm]{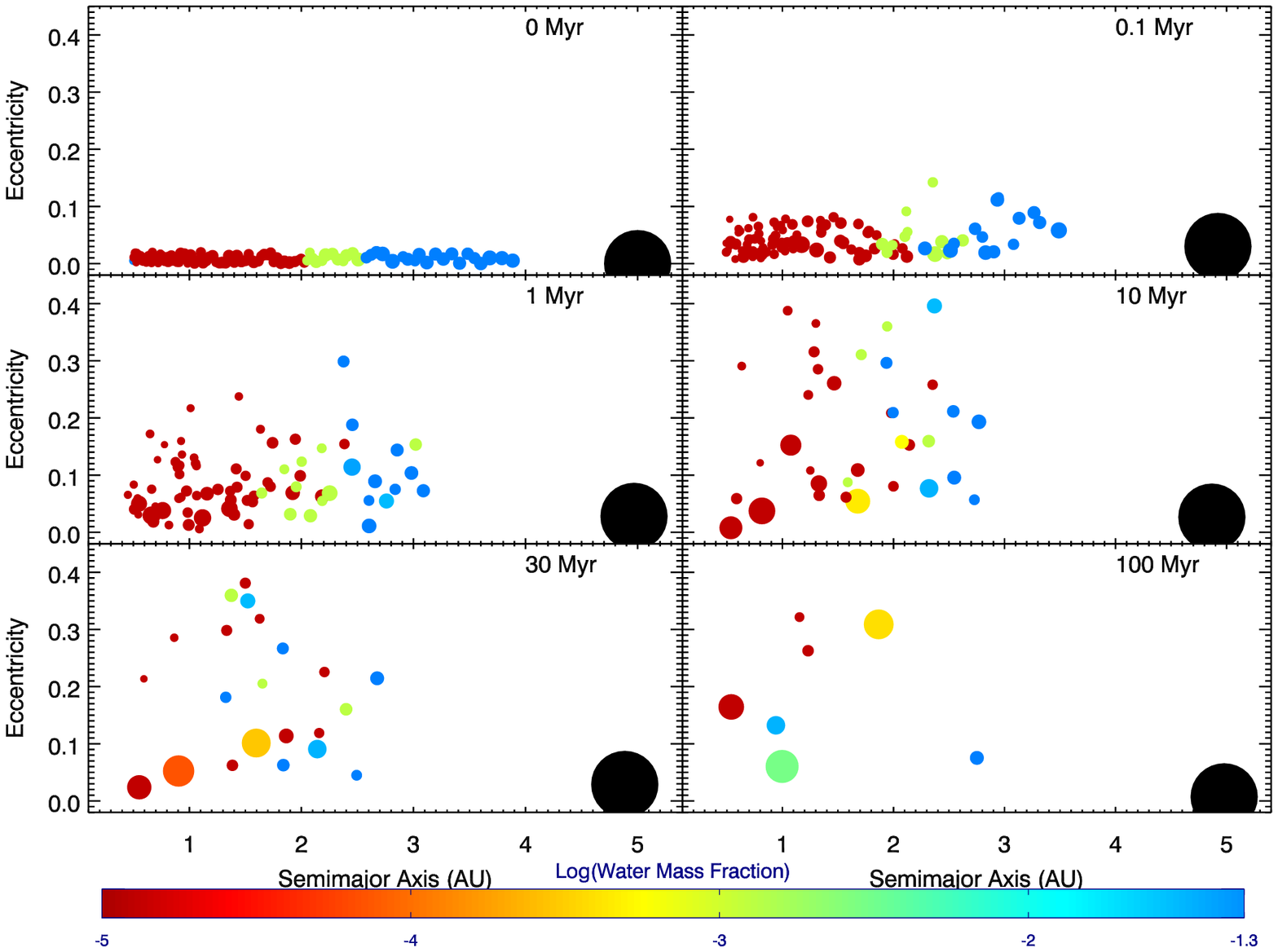}
\vskip 1in
\caption{Formation of Earth-like planets in a binary-planetary 
systems. The top panel shows simulations in a binary with a
separation of 30 AU, eccentricity of 0.2, and stellar 
components of 1 solar-mass. As shown here, an Earth-like planet 
(1.17 Earth-masses) with a water to mass ratio of 0.00164, 
is formed in the habitable zone of the primary  at 1.16 AU,
with an eccentricity of 0.02. The bottom panel shows the
formation of an Earth-like object in a binary with a solar-mass
primary and a 1.5 solar-masses secondary. The separation
of the binary in this case is 30 AU, the mass of the
Earth-like planets is 0.95 Earth-masses and its water to
mass ratio is 0.00226. The semimajor axis of this planet and
its orbital eccentricity are equal to 0.99 AU and 0.07,
respectively. For the sake of comparison, Sun's habitable zone 
is approximately at 0.95-1.15 AU, 
Earth's orbital eccentricity is 0.017, and Earth's
water to mass ratio is $\sim$ 0.001 \citep{Hagh07}.}
\label{fig:25}
\end{figure}
\begin{figure}
\centering
\includegraphics[height=12cm]{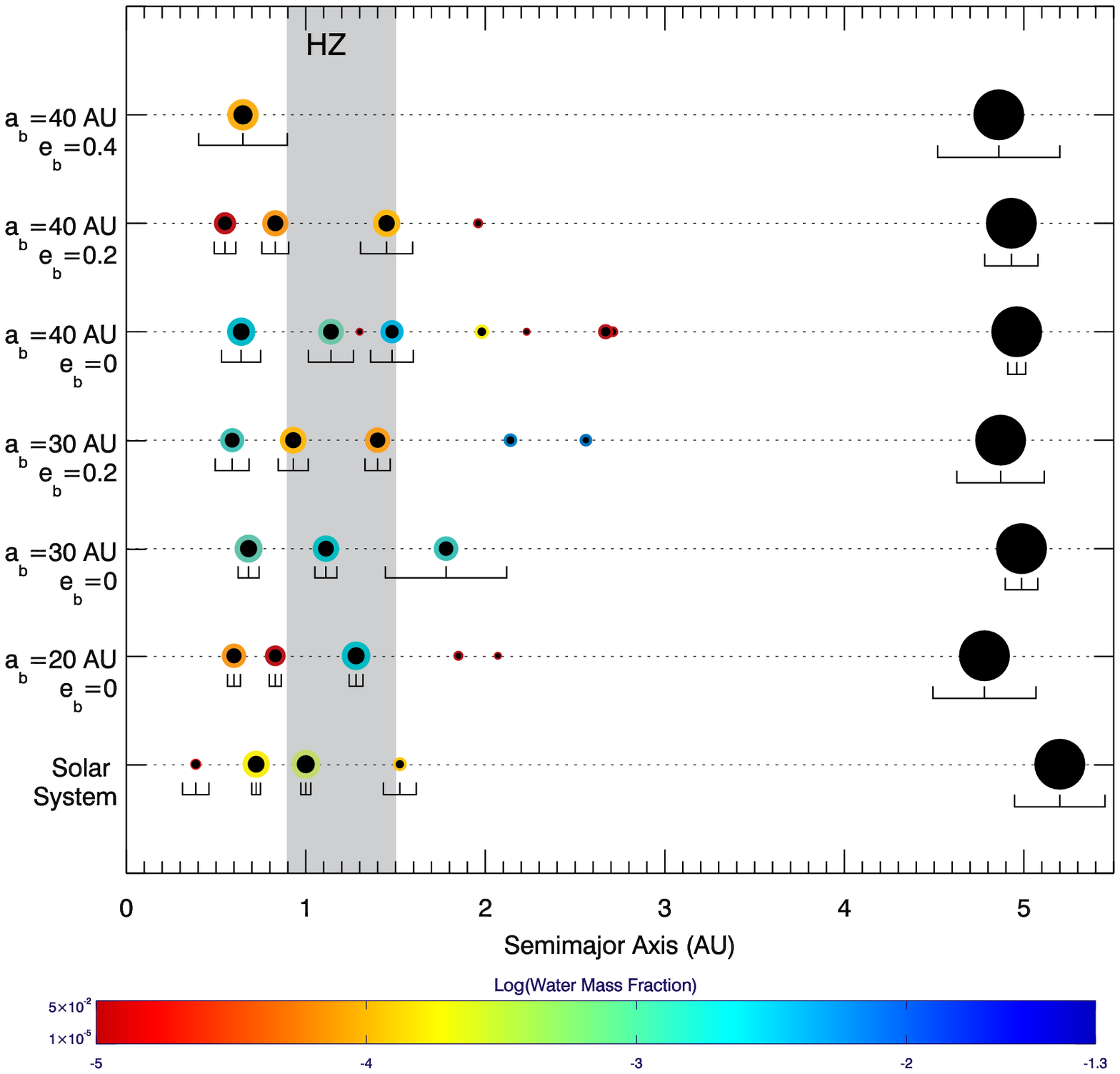}
\caption{Formation of Earth-like planets in different 
binary-planetary systems. As shown here, for a given binary mass-ratio,
the delivery of water to terrestrial region becomes less efficient
as the perihelion distance of the binary becomes smaller
\citep{Hagh07}.}
\label{fig:26}
\end{figure}
\noindent
can also
be applied to the dynamics of a terrestrial planet. 
Equation (9.1) and the
stability conditions presented by figure 9.5 can be used to
determine the long-term stability of an Earth-like object
in a binary system.

If a binary contains giant planets, however, 
the situation is different. The gravitational perturbations 
of the latter objects will
have significant effects on the motion and dynamics
of terrestrial planets in the system. 
As shown by Haghighipour (2006), an Earth-size object,
in a region between the giant planet and the primary of 
$\gamma$ Cephei binary system, can maintain its stability
\begin{figure}
\centering
\vskip 0.5in
\includegraphics[height=15cm]{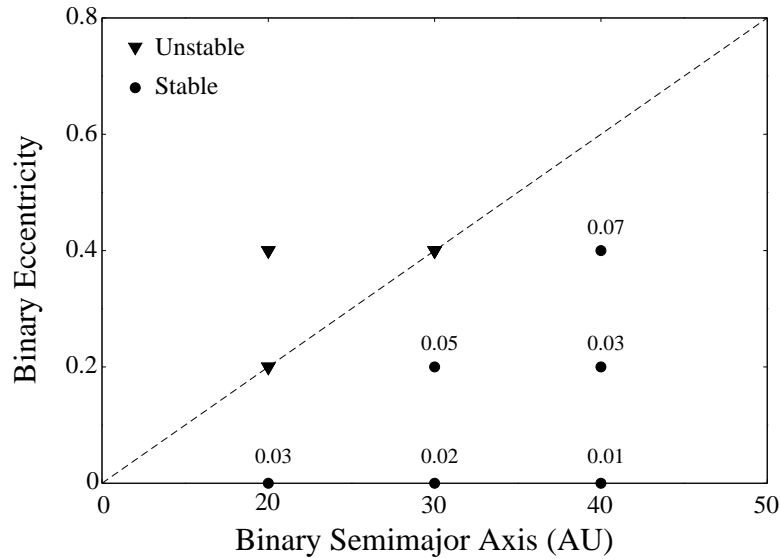}
\vskip -3in
\caption{The $({e_b},{a_b})$ parameter-space of an equal-mass 
binary-planetary system. Circles correspond to binaries with initial 
parameters chosen from figure 9.25, in which habitable planets were formed. 
Triangles represent systems in which the giant planet became unstable. 
The numbers associated with each circle represents the mean eccentricity of the
giant planet of the system at the end of the simulation. As shown here,
moderately close binaries with lower eccentricities (larger perihelia)
are more suitable places for the formation of habitable planets
\citep{Hagh07}.}
\label{fig:27}
\end{figure}
\noindent
only in orbits close to the primary star and outside
the influence zone\footnote{The influence zone
of a planetary object with a mass $m_p$ around a star
with a mass $M$ is defined as the region between
$3{R_H}-{a_p}(1-{e_p})$ and $3{R_H}+{a_p}(1+{e_p})$,
where ${a_p}$ is the semimajor axis
of the planet, and ${R_H}={a_p}{({m_p}/{3M})^{1/3}}$ is its Hill radius}
of the giant body. Integrating
the equations of motion of a full four-body system,
this author has shown that an Earth-like planet will
not be able to sustain a stable orbit in the habitable zone of
$\gamma$ Cephei's primary star (figure 9.24). 
However, it is possible for such an object to have 
a stable orbit when 
$0.3 \leq {a_T} \leq 0.8$ AU, ${0^\circ} \leq {i_T}=
{i_p} \leq {10^\circ}$, and ${e_b}\leq 0.4$.  Here $a_T$
represents the semimajor axis of the terrestrial planet 
and $i_T$ is its orbital inclinations 
with respect to the plane of the binary.

As mentioned above, the instability of an Earth-like planet in the 
habitable zone of $\gamma$ Cephei can be attributed to the interaction
between this object and the giant planet of the system. When the
Earth-like planet is outside the giant planet's influence zone 
(e.g., at closer distances to the primray star) it can maintain its orbit for
several hundred million years.
Figure 9.24 suggest that, in order for a binary-planetary
system to be habitable, its habitable zone has to be outside
the influence region of its giant planet. In an S-type binary-planetary
system, this implies a primary with a close-in habitable region.
In a recent article, \citet{Hagh07} have studies
the habitability of such a system. By considering a binary with a
Sun-like primary star and a Jupiter-sized planet in a circular orbit
at 5 AU, and by adopting the model of \citet{Morbidelli00}, 
which is based on
the assumption that water-carrying objects, in the Sun's asteroid
belt, were the primary source of the delivery of water to Earth,
these authors integrated the orbits of a few hundred
protoplanetary (Moon- to Mars-sized) objects, and showed that it
is indeed possible to form Earth-sized planets, with substantial
amount of water, in the habitable zone of the primary star
(figure 9.25). As shown by these authors, the mass and orbital
parameters of the secondary star play important roles in
the radial mixing of protoplanetary objects and the delivery of water
to the habitable zone of the primary star. The giant planet of the
system also plays the important role of transferring angular momentum
from the secondary star to the disk of protoplanets, and
enhancing the radial mixing of these objects. As shown in figure 9.26,
water delivery is less efficient in binaries with smaller perihelia
since in such systems, the close approach of the binary companion to
the giant planet increases its eccentricity, which in turn results in
stronger interaction between this object and the disk of protoplanets,
causing them to become unstable in very short time. The results of
the simulations by \citet{Hagh07} indicate that 
binary-planetary systems with giant planets at  5-10 AU, and
binary perihelion distances of approximately 20 AU to 25 AU, 
will be more efficient in forming and 
hosting habitable planets (figure 9.27).

\section {Future Prospects}

The discovery of planets in binary star systems is
one the interesting surprises of modern astronomy. Despite the long 
history of the study of planets in such environments, the recent detection of
planets in moderately close binaries has 
confronted astronomers with many new challenges. Many aspects of
the formation process of these planets are still
unresolved, and questions regarding their frequency and detection
techniques demand more detailed investigation.

The habitability of binary systems is also an open question.
Although recent simulations of the late stage of terrestrial
planet formation in binary systems have indicated that water-carrying
planets can form in the habitable zone of a binary-planetary system,
more studies are necessary to understand how protoplanetary objects
can develop and evolve in such an environment.
Such studies have implications for 
investigating the habitability of extrasolar planets, and 
tie directly to several of near-future NASA missions, 
in particular, the space mission Kepler.

\clearpage

\begin{figure}
\vskip 2in
\centering
\includegraphics[height=9cm]{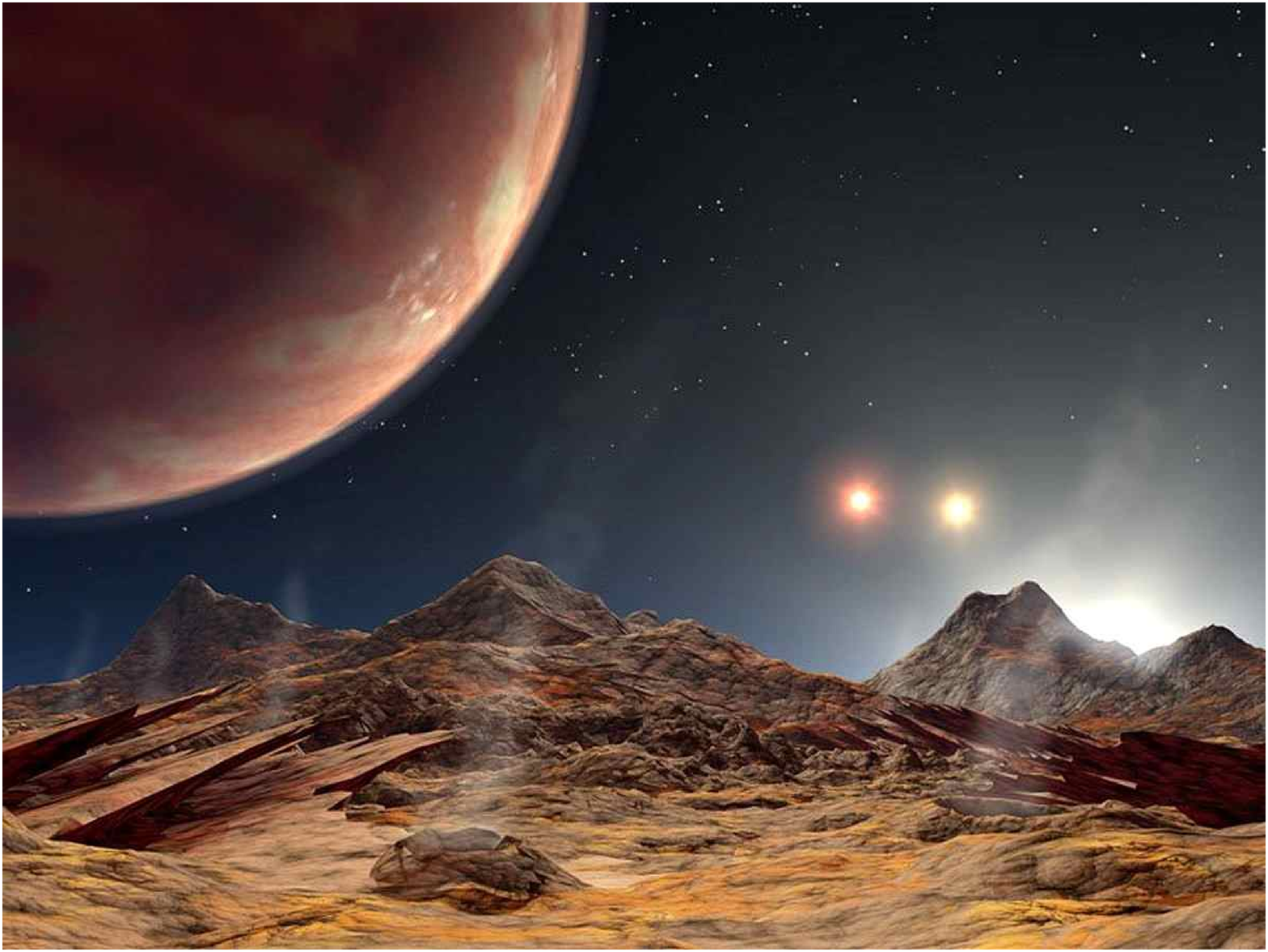}
\caption{Artistic rendition of the view from the moon of
a giant planet in a triple star system. Figure from JPL-Caltech/NASA.}
\label{fig:0}
\end{figure}

\end{document}